\documentclass[12pt,a4paper]{article}
\usepackage[latin1]{inputenc}
\usepackage{amssymb,amsmath,mathrsfs,bm}
\usepackage{comment}
\usepackage{booktabs}
\usepackage{mdframed}
\usepackage{graphicx}
\usepackage{setspace}
\usepackage{a4wide}
\usepackage{tcolorbox}
\usepackage{cite}
\usepackage{hyperref}
\usepackage{enumitem}
\usepackage{titlesec}

\usepackage{tikz}
\usetikzlibrary{calc}
\usepackage{color}

\makeatletter 
\@addtoreset{equation}{section}
\makeatother

\hypersetup{
  linktocpage,
  colorlinks  = true, 
  urlcolor    = blue, 
  linkcolor   = blue, 
  citecolor   = blue 
}

\def \be  {\begin{equation}}
\def \ee  {\end{equation}}
\def \ba  {\begin{eqnarray}}
\def \ea  {\end{eqnarray}}

\def \Tr {\mathop{\rm Tr}\nolimits}

\newcommand \Li{{\rm Li}}

\newcommand \widebar [1] {\overline{#1}}

\def\xb{\bar{x}}
\def\dbar#1{\widebar{D}_{#1}}

\def\f{f_2}
\def\g{f_3}
\def\h{f_4}

\def\ap{\alpha'}

\def\dfour{\Delta^{(4)}}

\usepackage{tocloft}
\begin{document}

\usetikzlibrary{arrows}
\thispagestyle{empty}

~\\[-2.25cm]

\begin{center}
\vskip 0.5truecm 
{\Large\bf
%
{\LARGE Genus-one open string amplitudes \\\vskip.05in
	on AdS$_5\times$S$^3$ from CFT}
}\\
\vskip 1.25truecm
	{\bf H. Paul$^{1,2}$ and M. Santagata$^{3}$ \\
	}
		\vskip 0.4truecm
	{\it
		$^{{1})}$ Universit\'e Paris-Saclay, CNRS, CEA, Institut de Physique Th\'eorique, \\
		91191, Gif-sur-Yvette, France\\\vskip .2truecm
		$^{{2})}$ Instituut voor Theoretische Fysica, KU Leuven, Celestijnenlaan 200D, \\
		B-3001 Leuven, Belgium\\\vskip .2truecm
		$^{{3})}$ Department of Physics, National Taiwan University, Taipei 10617, Taiwan\\
		\vskip .2truecm   }
	\vskip .2truecm
\end{center}

\textit{E-mail:}\,  \href{mailto:hynek.paul@kuleuven.be}{{\tt hynek.paul@kuleuven.be}}, \, \href{mailto:michelesa@ntu.edu.tw}{{\tt michelesa@ntu.edu.tw}}

\vskip 1.25truecm

\centerline{\bf Abstract}
\vskip .4truecm
We bootstrap one-loop string corrections to the four-point function of half-BPS operators in a 4d $\mathcal{N}=2$ SCFT with flavour group $SO(8)$, dual to gluon scattering at genus one on AdS$_5\times$S$^3$. We identify an 8-dimensional organising principle which governs the spectrum of double-trace anomalous dimensions, valid to all orders in the string length. This has precise implications for the structure of one-loop Mellin amplitudes, which we explicitly compute for the first three orders beyond the field-theory limit. We also consider the corresponding position space representation, which is entirely determined by the square of a certain differential operator acting on a simpler ``pre-correlator''. Finally, we show that the flat-space limit of the Mellin amplitudes exactly matches the logarithmic terms of the genus-one amplitude in 8-dimensional flat space, which we compute via a partial-wave analysis.
\newpage
\setcounter{page}{1}\setcounter{footnote}{0}
\setcounter{tocdepth}{2}
\tableofcontents
\newpage
\section{Introduction and summary of results}
The study of scattering amplitudes in AdS has received particular attention in the past few years. 
Considerable progress has been made in the study of four-point scattering in various AdS$\times$S backgrounds with a known dual CFT description,\footnote{See \cite{Bissi:2022mrs,Heslop:2022xgp} for recent reviews on the subject.} especially in cases where the background possesses a conformally flat metric.
In fact, there is by now a lot of evidence that, at least in the half-BPS sector and in the strongly coupled regime, such theories enjoy a higher-dimensional hidden conformal symmetry.\footnote{A manifestation of this symmetry has been found also in $\mathcal{N}=4$ SYM at weak coupling \cite{Caron-Huot:2021usw,Caron-Huot:2023wdh}.} At present, there are four theories for which this is known to be the case: 
AdS$_5 \times$S$^5$ \cite{Caron-Huot:2018kta}, AdS$_3 \times$S$^3$ \cite{Rastelli:2019gtj,Giusto:2020neo}, AdS$_5 \times$S$^3$ \cite{Alday:2021odx} and AdS$_2 \times$S$^2$ \cite{Abl:2021mxo}. 
While a formal proof and a complete understanding of the origin of the symmetry is still lacking, its existence dramatically simplifies the computation of the dual correlation functions. Thus, these theories provide an ideal playground to test our understanding of scattering amplitudes in AdS.

The most successful example is arguably the computation of four-point correlation functions of half-BPS operators in $\mathcal{N}=4$ SYM at strong coupling, dual to the scattering of four closed strings in AdS$_5 \times$S$^5$. By now, there is a plethora of available results, both in the supergravity limit and including string corrections, see e.g. \cite{Rastelli:2016nze,Aprile:2017bgs,Alday:2018pdi,Binder:2019jwn,Drummond:2019odu,Aprile:2019rep,Alday:2019nin,Drummond:2020dwr,Aprile:2020luw,Bissi:2020woe,Aprile:2020mus,Huang:2021xws,Drummond:2022dxw,Aprile:2022tzr,Alday:2023mvu,Brown:2023zbr} and references therein.

An analogous program has been recently carried out also for four-point scattering of supergluons in AdS$_5 \times$S$^3$, dual to half-BPS operators $\mathcal{O}_p$ of protected dimension $p$. 
This theory is an orientifold of type IIB string theory in which D7-branes are localised at the orientifold fixed planes with D3-branes, and can be engineered from F-theory on a $D_4$ singularity \cite{Sen:1996vd}.
The dynamics of the $N$ D3-branes is described by a certain $USp(2N)$ $\mathcal{N}=2$ SCFT with flavour group $SO(8)$. 
In this particular string-theory realisation, the dual field theory has a vanishing beta function and the parameters of two theories are related by
\begin{align}\label{eq:dictionary}
	\lambda=\frac{R^4}{\ap^2} = 8\pi g_s N\,,
\end{align}
where $\lambda=g^2_{\text{IR}}N$ with $g_{\text{IR}}$ being the renormalised Yang-Mills coupling in the IR. Since the theory is exactly superconformal, one can take advantage of CFT techniques to investigate its strong-coupling dynamics. In particular, as  found in \cite{Alday:2021odx}, the tree-level four-point dynamics in the gluon sector is controlled by a 8-dimensional hidden conformal symmetry, which suggests that, at least in this sector of the theory, observables should follow simple patterns, similar to those found in other AdS$\times$S-type backgrounds. These expectations have been confirmed by a number of recent results at various orders in $1/N$ and $1/\lambda$ \cite{Alday:2021ajh,Drummond:2022dxd,Huang:2023oxf,Glew:2023wik,Behan:2023fqq}.

In this paper we will continue the exploration of four-point correlators of operators $\mathcal{O}_2$ at large $N$ and $\lambda$. Using the holographic dictionary \eqref{eq:dictionary}, this corresponds to the genus and low-energy expansion in small $g_s$ and $\ap$ of the open string amplitude. Our main object of study is the reduced Mellin amplitude $\mathcal{M}^{I_1I_2I_3I_4}(s,t)$, whose precise definition will be given in Section \ref{sec:setup}. For now, let us just state that in the aforementioned limit the Mellin amplitude has the expansion
\begin{align}\label{eq:M_large_N}
	\mathcal{M} = \frac{1}{N}\mathcal{M}_{\text{gluon}}^{(1)} + \frac{1}{N^2}\big(\mathcal{M}^{(2)}_{\text{gluon}} + \mathcal{M}_{\text{grav}}^{(2)}\big) + O(1/N^3)\,,
\end{align}
where for simplicity we have suppressed all kinematic labels. The first term, $\mathcal{M}_{\text{gluon}}^{(1)}(s,t)$, is the contribution from tree-level gluon exchanges, and corresponds to an AdS version of the Veneziano amplitude, while $\mathcal{M}_{\text{gluon}}^{(2)}(s,t)$ corresponds to the genus-one string amplitude. Note that at order $1/N^2$ there is an additional contribution coming from tree-level graviton exchanges.\footnote{
The tree-level graviton contribution has been discussed in \cite{Alday:2021ajh}, see their Section 6. 
} However, as we will see, such tree-level terms will not interfere in our construction of one-loop amplitudes, and we can therefore focus only on the gluon contributions in this work. We will henceforth drop the ``gluon'' subscript.

Each of the above terms admits a further expansion at large $\lambda$ (or, analogously small string length $\alpha'$). For the tree-level contribution $\mathcal{M}^{(1)}$ this reads
\begin{align}\label{eq:M1_lambda_exp}
	\mathcal{M}^{(1)} = \mathcal{M}^{(1,0)} + \lambda^{-1}\mathcal{M}^{(1,2)} + \lambda^{-\frac{3}{2}}\mathcal{M}^{(1,3)} + \lambda^{-2}\mathcal{M}^{(1,4)} + O(\lambda^{-\frac{5}{2}})\,.
\end{align}
The leading term $\mathcal{M}^{(1,0)}$ is the field-theory contribution, addressed in \cite{Alday:2021odx}, which is followed by an infinite tower of higher-derivative corrections $\mathcal{M}^{(1,m\geq2)}$, which have been recently considered in \cite{Glew:2023wik,Behan:2023fqq}. A brief review of these results is given in Section \ref{sec:tree_level_review}, and in more detail in Appendix \ref{app:corrgencharges}, where we review the generalisation to correlators of arbitrary external charges.

Next, at order $1/N^2$, the one-loop gluon amplitude from \eqref{eq:M_large_N} has the expansion
\begin{align}
	\mathcal{M}^{(2)} = \mathcal{M}^{(2,0)} + \log(\lambda)\,\mathcal{M}_{\log} + \lambda^{-1}\mathcal{M}^{(2,2)} + \lambda^{-\frac{3}{2}}\mathcal{M}^{(2,3)} + \lambda^{-2}\mathcal{M}^{(2,4)} + O(\lambda^{-\frac{5}{2}})\,,
\end{align}
with the first term being the one-loop field theory correlator $\mathcal{M}^{(2,0)}$, computed in \cite{Alday:2021ajh} (up to a contact term ambiguity corresponding to a genus-one correction of the tree-level $\mathcal{M}^{(1,2)}$ term). The divergence of the one-loop term is regularised by a logarithmic contribution $\mathcal{M}_{\log}$ derived in \cite{Behan:2023fqq}. The goal of this work is to address the first few one-loop string corrections $\mathcal{M}^{(2,m)}$, $m=2,3,4$, which are explicitly given in Section \ref{sec:oneloopamp}. As we will better explain later on, the 8-dimensional hidden symmetry allows to drastically simplify the computation, especially at the first few orders: it turns out that the leading logs are given by application of a certain differential operator, denoted by $\dfour$, on the tree-level discontinuity. We then also consider the corresponding position space representation $\mathcal{H}^{(2,m)}$ of these string amplitudes. Notably, we find that the differential operator mentioned above greatly simplifies their presentation. In particular, we have
\begin{align}
	\mathcal{H}^{(2,m)} = \big(\dfour\big)^2 \mathcal{P}^{(2,m)}\,,
\end{align}
for some ``pre-correlators'' $\mathcal{P}^{(2,m)}$, which are considerably simpler than the full correlator.

The remainder of this paper is organised as follows. In \textbf{Section \ref{sec:setup}}, we discuss some prerequisites and introduce the decomposition into flavour channels and superconformal blocks. \textbf{Section \ref{sec:leading_log}} is about the construction of the one-loop leading logs: after discussing the OPE predictions, we notice how the 8-dimensional hidden symmetry organises the string-corrected double-trace spectrum. This in turn results in compact formulae for the leading logs. A comment about the general colour structure of loop-amplitudes is given in Section \ref{sec:colour_structure_loops}. In \textbf{Section \ref{sec:oneloopamp}}, we explicitly construct the one-loop amplitudes up to order $\lambda^{-2}$, both in Mellin and in position space. In \textbf{Section \ref{sec:flat-space-limit}}, we show that the flat-space limit of our results is in agreement with the discontinuity of the one-loop amplitude, which we obtain through a partial-wave analysis of the 8-dimensional flat-space amplitude. This provides a consistency check of our one-loop computations. Finally, in \textbf{Section \ref{sec:final}}, we conclude and outline some future directions. 

\section{Setup}\label{sec:setup}
In this paper we will study four-point function of half-BPS operators in a certain $USp(2N)$ $\mathcal{N}=2$ SCFT with flavour group\footnote{This is a gauge group from a bulk perspective, as dictated by the AdS/CFT correspondence.} $SO(8)$, dual to gluon scattering on a AdS$_5 \times$S$^3$ background.
The half-BPS operators we are interested in are of the form $\mathcal{O}_p^{I; a_1,\ldots, a_p; \bar{a}_1,\ldots, \bar{a}_{p-2}}$. These operators have protected dimension $\Delta=p=2,3,\ldots$, they are chargeless under $U(1)_R$, and transform in the adjoint of $SO(8)$. Here $I$ is the colour index, $a_1,\ldots, a_p$ are symmetrised $SU(2)_R$ R-symmetry indices and similarly $\bar{a}_i$ are indices of an additional $SU(2)_L$ flavour group, such that the above operator transforms in the spin-$\frac{p}{2}$ and spin-$\frac{p-2}{2}$ representations of $SU(2)_R$ and $SU(2)_L$. A convenient way to deal with these various indices is by contracting them with auxiliary bosonic two-component vectors $\eta$ and $\bar{\eta}$:
\begin{equation}\label{contractedO}
	\mathcal{O}_p^{I}(x;\eta,\bar\eta) \equiv \mathcal{O}_p^{I; a_1,\ldots, a_p; \bar{a}_1,\ldots, \bar{a}_{p-2}}(x)\,\eta_{a_1} \cdots \eta_{a_p}\, \bar{\eta}_{\bar{a}_1} \cdots \bar{\eta}_{\bar{a}_{p-2}}\,.
\end{equation}
Note the operator with lowest dimension, i.e. $p=2$, transforms trivially under $SU(2)_L$:
\begin{equation}\label{contractedO2}
	\mathcal{O}_2^{I}(x;\eta) = \mathcal{O}_2^{I;a_1 a_2}(x)\,\eta_{a_1} \eta_{a_2}\,.
\end{equation}

\subsection{Four-point functions of supergluons}
The main subject of the paper will be the four-point function of the aforementioned half-BPS operators, which we will denote by
\begin{equation}
	G_{\vec{p}}^{I_1 I_2 I_3 I_4}(x_i,\eta_i, \bar{\eta}_i)\equiv \langle \mathcal{O}_{p_1}^{I_1}(x_1;\eta_1,\bar\eta_1) \mathcal{O}_{p_2}^{I_2}(x_2;\eta_2,\bar\eta_2) \mathcal{O}_{p_3}^{I_3}(x_3;\eta_3,\bar\eta_3) \mathcal{O}_{p_4}^{I_4}(x_4;\eta_4,\bar\eta_4) \rangle\,.
\end{equation}
Due to the definition \eqref{contractedO}, the dependence on the variables $\eta_i$, $\bar\eta_i$ is polynomial with its degree dictated by the external charges $p_i$. We can further exploit the conformal symmetries to write the above correlator as a function of cross-ratios only. Specialising to the case of interest, i.e. the simplest correlator with equal dimensions $p_i=2$, we have\footnote{We review the analogous formula for arbitrary charges in appendix \ref{app:corrgencharges}, with the conventions as in \cite{Drummond:2022dxd}.}
\begin{align}\label{eq:G_2222}
	G_{2222}^{I_1 I_2 I_3 I_4}(x_i,\eta_i,\bar\eta_i) = \frac{\langle\eta_1\eta_3\rangle^2\langle\eta_2\eta_4\rangle^2}{(x_{13}^2x_{24}^2)^2}\,\mathcal{G}^{I_1 I_2 I_3 I_4}(U,V;y)\,,
\end{align}
where the function $\mathcal{G}$ is a polynomial of degree 2 in $y$, the $\eta$-variables are contracted via $\langle \eta_i \eta_j\rangle = \eta_{i}^a \eta_{j}^b \epsilon_{ab}$, and we define the cross-ratios as
\begin{align}
	U=x\xb=\frac{x_{12}^2x_{34}^2}{x_{13}^2 x_{24}^2}\,,\quad V=(1-x)(1-\xb)=\frac{x_{14}^2x_{23}^2}{x_{13}^2 x_{24}^2}\,,\quad y=\frac{\langle \eta_1 \eta_2 \rangle \langle \eta_3 \eta_4 \rangle}{\langle \eta_1 \eta_3 \rangle\langle \eta_2 \eta_4 \rangle}\,.
\end{align}
So far we have used only the bosonic symmetries. Further constraints can be obtained by considering the fermionic generators of the superconformal group, leading to the so-called superconformal Ward identities \cite{Nirschl:2004pa}. The solution to these additional constraints takes the form
\begin{align}\label{eq:solution_SCWI}
	\mathcal{G}^{I_1 I_2 I_3 I_4}(U,V;y) = \mathcal{G}_0^{I_1 I_2 I_3 I_4}(U,V;y)+\mathcal{I}\,\mathcal{H}^{I_1 I_2 I_3 I_4}(U,V)\,,\quad\mathcal{I}=(x-y)(\xb-y)\,.
\end{align}Importantly, the \textit{reduced} correlator $\mathcal{H}^{I_1 I_2 I_3 I_4}(U,V)$ contains all the coupling dependence of the correlator and, as indicated, no longer depends on the $SU(2)_R$ R-symmetry cross-ratio $y$. On the other hand, the \textit{protected} part $\mathcal{G}_0^{I_1 I_2 I_3 I_4}(U,V;y)$ is coupling-independent. As we will see later, we only need to know its leading order large $N$ contribution, which corresponds to disconnected contributions to the correlator and can be computed in generalised free field theory. However, note that the above splitting \eqref{eq:solution_SCWI} is in general \textit{not} unique, a fact which will become important in Section \ref{sec:block_decomposition} when we introduce the superconformal block decomposition. For now, we define $\mathcal{G}_0$ and $\mathcal{H}$ by requiring that each of them is crossing symmetric by itself.

\paragraph{Flavour symmetry:} In order to discuss the properties under crossing, we need to take care of one additional complication which correlators describing scattering of gluons exhibit, namely the flavour indices. In the present case, the gauge group is given by $SO(8)$ and as noted above the operators $\mathcal{O}_p^{I}$ transform in the adjoint irrep $\mathbf{28}$. A convenient way to process the flavour symmetry is to decompose the correlator into irreps -- labelled by $\mathbf{a}$ -- which appear in the tensor product of the adjoint representation with itself:
\begin{align}\label{eq:adjxadj}
	\mathbf{a}\in\mathbf{28}\otimes\mathbf{28}=\underbrace{\mathbf{1}\oplus\mathbf{35_v}\oplus\mathbf{35_c}\oplus\mathbf{35_s}\oplus\mathbf{300}}_{\text{symmetric}}\oplus\underbrace{\mathbf{28}\oplus\mathbf{350}}_{\text{antisymmetric}},
\end{align}
where we ordered the irreps (or flavour channels) according to their parity, i.e. their symmetry under $1\leftrightarrow2$ exchange. Such a decomposition of the correlator is achieved by introducing projection operators $P_\mathbf{a}^{I_1I_2I_3I_4}$, whose job is to project the external flavour indices onto the above irreps. For the case of $SO(8)$, these projectors have been constructed in e.g. \cite{Isaev:2020kwc}.\footnote{In our implementation, we have found it useful to employ the ``birdtrack'' notation as described in \cite{cvitanovic2008group}.}

Before proceeding, let us mention two important properties of the projectors. Firstly, being properly normalised projection operators, they are idempotent and taking the trace computes the dimension of the irrep they project onto:
\begin{align}
	P_\mathbf{a}^{I_1I_2I_3I_4}P_\mathbf{b}^{I_4I_3I_5I_6}=\delta_{\mathbf{a},\mathbf{b}}P_\mathbf{b}^{I_1I_2I_5I_6},\quad \Tr(P_\mathbf{a})=P_\mathbf{a}^{I_1I_2I_2I_1}=\text{dim}(\mathbf{a})\,.
\end{align} 
Secondly, when considering crossing transformations of the correlator, the flavour indices get permuted accordingly. We thus need to compare the projectors with permuted indices to the original ones. This simply corresponds to a change of basis, whose action is encoded in a crossing matrix. For instance, swapping positions $1\leftrightarrow2$ acts diagonally and simply measures the parity of the irreps $\mathbf{a}$. From \eqref{eq:adjxadj}, we therefore find that the $s$-channel crossing matrix is diagonal with entries
\begin{align}\label{eq:s_crossing_mat}
	F_s = \text{diag}\big(1,1,1,1,1,-1,-1\big).
\end{align}
On the other hand, the $t$- and $u$-channel crossing matrices -- corresponding to exchanging operators at positions $1\leftrightarrow3$ and $1\leftrightarrow4$, respectively -- necessitate a computation. They are computed by
\begin{align}
	(F_t)_\mathbf{a}^{~\mathbf{b}}=\frac{1}{\text{dim}(\mathbf{a})}P_\mathbf{a}^{I_1I_2I_3I_4}P_\mathbf{b}^{I_3I_2I_1I_4}\,,\quad (F_u)_\mathbf{a}^{~\mathbf{b}}=\frac{1}{\text{dim}(\mathbf{a})}P_\mathbf{a}^{I_1I_2I_3I_4}P_\mathbf{b}^{I_4I_2I_3I_1}\,,
\end{align}
and the explicit expressions for these matrices are recorded in Appendix \ref{app:colour}.

Finally, applying this decomposition into $SO(8)$ flavour channels to the reduced correlator, we have
\begin{align}\label{eq:colour_decomp_H}
	\mathcal{H}^{I_1I_2I_3I_4}(U,V) = \sum_{\mathbf{a}\in\mathbf{28}\otimes\mathbf{28}}\mathcal{H}_\mathbf{a}(U,V)\,P_\mathbf{a}^{I_1I_2I_3I_4}\,,
\end{align}
which effectively decouples the flavour structure from the dynamical information of the correlator. After this projection onto irreps, it is useful to think of $\mathcal{H}_\mathbf{a}(U,V)$ as a vector in colour space with components ordered as in \eqref{eq:adjxadj}:
\begin{equation}\label{eq:vector_notation}
\mathcal{H}_\mathbf{a}(U,V) = \begin{pmatrix} \mathcal{H}_\mathbf{1}(U,V)\\ \mathcal{H}_\mathbf{35_v}(U,V)\\ \mathcal{H}_\mathbf{35_c}(U,V)\\ \mathcal{H}_\mathbf{35_s}(U,V)\\ \mathcal{H}_\mathbf{300}(U,V)\\\hline \mathcal{H}_\mathbf{28}(U,V)\\ \mathcal{H}_\mathbf{350}(U,V) \end{pmatrix},
\end{equation}
where the horizontal line is nothing but a guide to the eye, separating the symmetric from antisymmetric irreps.

With this in place, the full crossing symmetry of the correlator \eqref{eq:G_2222} implies the relations
\begin{align}\label{eq:crossing_H}
	\mathcal{H}_\mathbf{a}(U,V) = \frac{1}{V^3}(F_s)_\mathbf{a}^{~\mathbf{b}}\, \mathcal{H}_\mathbf{b}(U/V,1/V) = (F_t)_\mathbf{a}^{~\mathbf{b}}\, \mathcal{H}_\mathbf{b}(V,U) = \frac{1}{U^3}(F_u)_\mathbf{a}^{~\mathbf{b}}\,\mathcal{H}_\mathbf{b}(1/U,V/U)\,,
\end{align}
among the seven different flavour channels of \eqref{eq:vector_notation}.

\paragraph{Mellin space:} In the context of holographic correlators, it has often turned out to be beneficial to consider the Mellin transform of the correlator. For our purposes, it is useful to work directly with the so-called (reduced) Mellin amplitude $\mathcal{M}^{I_1I_2I_3I_4}(s,t)$, defined in terms of the reduced correlator by
\begin{equation}\label{eq:Mellin_trafo_2222}
	\mathcal{H}^{I_1I_2I_3I_4}(U,V) = \int_{-i\infty}^{i\infty} \frac{ds\, dt}{(2\pi i)^2} \, U^s V^t \,\mathcal{M}^{I_1I_2I_3I_4}(s,t)\,\Gamma^2(-s)\Gamma^2(-t)\Gamma^2(-u)\,,
\end{equation}
where the Mellin variables $s,t,u$ obey the constraint equation
\begin{align}
	s+t+u=-3\,.
\end{align}
This formulation has the advantage that, at tree-level, all contributions of double-trace operators are naturally encoded in the gamma functions present in \eqref{eq:Mellin_trafo_2222}. As such, poles in the Mellin amplitude correspond to exchanges of single-trace operators. These are known to be absent for the tree-level string corrections $\mathcal{M}^{(1,m\geq2)}$, which are therefore simply polynomials of the Mellin variables. This is no longer true at one-loop order, where the analytic structure predicted by the OPE decomposition -- which we introduce in the next section -- allows for additional simple poles at the double-trace locations.

Lastly, crossing transformations act by simply permuting the Mellin variables. Using an analogous colour decomposition to \eqref{eq:colour_decomp_H} for the Mellin amplitude, $\mathcal{M}^{I_1I_2I_3I_4}(s,t)=\mathcal{M}_\mathbf{a}(s,t)P^{I_1I_2I_3I_4}_\mathbf{a}$, crossing symmetry implies 
\begin{align}\label{eq:crossing_M}
	\mathcal{M}_\mathbf{a}(s,t) = (F_s)_\mathbf{a}^{~\mathbf{b}}\, \mathcal{M}_\mathbf{b}(s,u) = (F_t)_\mathbf{a}^{~\mathbf{b}}\, \mathcal{M}_\mathbf{b}(t,s) = (F_u)_\mathbf{a}^{~\mathbf{b}}\, \mathcal{M}_\mathbf{b}(u,t)\,.
\end{align}

\subsection{Superconformal block decomposition}\label{sec:block_decomposition}
Another crucial ingredient to our one-loop bootstrap program is the superconformal block decomposition of the correlator $\mathcal{G}^{I_1I_2I_3I_4}(U,V;y)$, which we briefly review here. To this end, it is useful to rewrite the previously given solution to the superconformal Ward identities, \eqref{eq:solution_SCWI}, in the form \cite{Nirschl:2004pa}
\begin{align}\label{eq:solution_SCWI_2}
	\mathcal{G}(U,V;y)=\frac{(y-\xb)x f(\xb)-(y-x)\bar{x} f(x)}{y(x-\xb)}+\mathcal{I}\,\mathcal{K}(x,\xb)\,,
\end{align}
where we recall the definition $\mathcal{I}=(x-y)(\xb-y)$. Splitting the correlator this way into a single-variable function $f(x)$ and a genuine two-variable contribution $\mathcal{K}(x,\xb)$ allows one to isolate the contributions from unprotected, long multiplets as they contribute only to $\mathcal{K}$ and not to $f$. We will therefore refer to $\mathcal{K}$ as the \textit{long} part of the correlator. Compared to \eqref{eq:solution_SCWI}, this simply amounts to a reshuffling of certain terms from the free correlator $\mathcal{G}_0$ into $\mathcal{K}$. Note that, in contrast to the reduced correlator $\mathcal{H}$ defined by \eqref{eq:solution_SCWI}, the long part is not crossing symmetric by itself -- instead, $f$ and $\mathcal{K}$ mix under crossing.

For what follows, we will be only interested in the long contributions and we thus restrict our attention to the function $\mathcal{K}(x,\xb)$. 
Before proceeding with the details of the decomposition into (super)conformal blocks, let us note that this decomposition is completely orthogonal to the previously introduced colour decomposition. In fact, for notational simplicity we have suppressed all colour indices in \eqref{eq:solution_SCWI_2}, as they present an additional structure on top of the superconformal properties. Nevertheless, it is useful to take care of the flavour indices by decomposing into $SO(8)$ irreps as in \eqref{eq:colour_decomp_H}. The flavour channels of the long part $\mathcal{K}_\mathbf{a}(x,\xb)$ then admit a conformal block decomposition according to
\begin{align}\label{eq:block_deco}
	\mathcal{K}_\mathbf{a}(x,\xb) = \sum_{\tau,\,\ell}A_{\mathbf{a},\tau,\ell}\,\mathcal{B}_{\tau,\ell}(x,\xb)\,,
\end{align}
where the above sum is over all exchanged long superconformal primaries $\mathcal{O}_{\tau,\ell}$, which we label by their twist $\tau=\Delta-\ell$ and spin $\ell$. Since each flavour channel has a definite parity under $1\leftrightarrow2$ exchange, c.f. equation \eqref{eq:adjxadj}, the sum over spins runs only over even (odd) values of $\ell$ for symmetric (antisymmetric) irreps $\mathbf{a}$. The OPE coefficients $A_{\mathbf{a},\tau,\ell}$ denote the squared three-point functions of the two external $p=2$ half-BPS operators and the exchanged operator, $\langle\mathcal{O}_2\mathcal{O}_2\mathcal{O}_{\tau,\ell}\rangle^2\vert_\mathbf{a}$. Finally, the long blocks $\mathcal{B}_{\tau,\ell}(x,\xb)$ are related to the standard four-dimensional conformal block by a shift of 2 in the twist. Explicitly, they are given by\footnote{Here we only quote the blocks applicable to the correlator with external charges $p_i=2$. For the case of arbitrary external charges, the long blocks are given by the product of the conformal and an additional ``internal block'' which takes care of the non-trivial $SU(2)_L\times SU(2)_R$ representations. These general blocks can be found in \cite{Drummond:2022dxd}, and a brief summary is presented in Appendix \ref{app:long_blocks}.}
\begin{align}\label{eq:long_block}
	\mathcal{B}_{\tau,\ell}(x,\xb) = \frac{(-1)^\ell}{U^2(x-\xb)}\Big[ \mathcal{F}_{\tfrac{\tau}{2}+1+l}(x)\,\mathcal{F}_{\tfrac{\tau}{2}}(\bar{x})-(x\leftrightarrow\xb)\Big],
\end{align}
with
\begin{align}
	\mathcal{F}_h(x)=x^h\,{}_2F_1(h,h,2h;x)\,.
\end{align}
Note that as a consequence of the hypergeometric differential equation, the long blocks satisfy an eigenvalue equation:
\begin{align}\label{eq:eigenvalue_blocks}
	\dfour\,\mathcal{B}_{\tau,l} = \delta_{\tau,l}^{(4)}\,\mathcal{B}_{\tau,l}\,,
\end{align}
where the differential operator $\dfour$ is given by
\begin{align}
	\dfour=\frac{1}{U^2(x-\xb)} \, D_xD_{\xb} \, U^2(x-\xb)\,, \quad D_{x}= x^2 \partial_x(1-x)\partial_x\,,
\end{align}
and $D_{x}$ is such that
\begin{equation}
D_{x}\,\mathcal{F}_h (x) = h(h-1)\,\mathcal{F}_h (x)\,.
\end{equation}
The eigenvalue $\delta_{\tau,\ell}^{(4)}$ is a polynomial in twist $\tau$ and spin $\ell$, and takes the form\footnote{In the general case, it also depends on the $SU(2)_R$ quantum number, see \cite{Drummond:2022dxd}.}
\begin{equation}\label{eq:eigenvalue_d4}
	\delta_{\tau,\ell}^{(4)}=\left( \frac{\tau}{2}-1 \right) \frac{\tau}{2}\left( \frac{\tau}{2}+\ell \right) \left( \frac{\tau}{2}+\ell +1\right).
\end{equation}

Let us now comment on the spectrum of unprotected operators exchanged in the sum \eqref{eq:block_deco}. As mentioned in the introduction, we consider the large $N$, large $\lambda$ expansion of the correlator. In this limit, all unprotected single-trace operators (corresponding to stringy states in the bulk) are expected to decouple from the spectrum, and hence the remaining exchanged states are double-trace operators\footnote{Triple- and all other multi-trace operators will also contribute, but their contributions are further suppressed by powers of $1/N$.} constructed from products of two half-BPS operators. As such, their classical twist $\tau^{(0)}$ is quantised and takes the values $\tau^{(0)}=2n$ with $n=1,2,3,\ldots$. However, generically there are many such double-trace operators with the same (classical) quantum numbers. In fact, for a given twist $\tau^{(0)}$ and spin $\ell$ one can construct $n-1$ degenerate operators, which are of the schematic form 
\begin{align}\label{eq:double-trace_ops}
	\mathcal{O}_2 \square^{n-2} \partial^\ell \mathcal{O}_2\vert_{[00]}\,,\,\mathcal{O}_3 \square^{n-3} \partial^\ell \mathcal{O}_3\vert_{[00]}\,,\,\ldots\,,\,\mathcal{O}_p  \partial^\ell \mathcal{O}_p\vert_{[00]}\,,
\end{align}
where for simplicity we have dropped the flavour indices, and the notation $\mathcal{O}\vert_{[00]}$ stands for the projection onto the singlet of $SU(2)_L\times SU(2)_R$. The true exchanged eigenstates in the block decomposition \eqref{eq:block_deco} are then linear combinations of the double-trace operators listed above. At large $N$, the twists $\tau$ of these eigenstates and their OPE coefficients $A_{\mathbf{a},n,\ell}$ admit the expansion
\begin{align}
\begin{split}
	\tau_{\mathbf{a},n,\ell} &= 2n + \frac{2}{N}\,\gamma^{(1)}_{\mathbf{a},n,\ell} + \frac{2}{N^2}\,\gamma^{(2)}_{\mathbf{a},n,\ell} + \ldots\,,\\
	A_{\mathbf{a},n,\ell} &= A^{(0)}_{\mathbf{a},n,\ell} + \frac{1}{N}\,A^{(1)}_{\mathbf{a},n,\ell} + \frac{1}{N^2}\,A^{(2)}_{\mathbf{a},n,\ell} + \ldots \,,
\end{split}
\end{align}
where it is understood that starting at order $1/N$ each contribution is also a function of $\lambda$, and will therefore acquire an associated strong coupling expansion in $1/\lambda$.

Upon insertion into the superconformal block expansion \eqref{eq:block_deco}, this gives rise to the large $N$ expansion of the long part $\mathcal{K}_\mathbf{a}(x,\xb)$ which takes the form
\begin{align}\label{eq:large_N_exp_K}
	\mathcal{K}_\mathbf{a}(x,\xb) = \mathcal{K}_\mathbf{a}^{(0)}(x,\xb) + \frac{1}{N}\,\mathcal{K}_\mathbf{a}^{(1)}(x,\xb) + \frac{1}{N^2}\,\mathcal{K}_\mathbf{a}^{(2)}(x,\xb) + \ldots\,.
\end{align}
Comparing to the expansion of the rhs of \eqref{eq:block_deco}, one finds that the various terms have the following block decompositions:
\begin{align}\label{eq:log(u)_stratification}
\begin{split}
	\mathcal{K}_\mathbf{a}^{(0)} = ~~&\log^0(U)\sum A^{(0)} \,\mathcal{B}_{n,\ell}(x,\xb)\,,\\[3pt]
	\mathcal{K}_\mathbf{a}^{(1)} = ~~&\log^1(U)\sum A^{(0)}\gamma^{(1)}\, \mathcal{B}_{n,\ell}(x,\xb)\\
						 +&\log^0(U)\sum \big[A^{(1)}+2A^{(0)}\gamma^{(1)}\partial_{\Delta}\big]\,\mathcal{B}_{n,\ell}(x,\xb)\,,\\[3pt]
	\mathcal{K}_\mathbf{a}^{(2)} = ~~&\log^2(U)\sum \tfrac{1}{2}A^{(0)}(\gamma^{(1)})^2 \,\mathcal{B}_{n,\ell}(x,\xb)\\
						 +&\log^1(U)\sum \big[A^{(1)}\gamma^{(1)}+A^{(0)}\gamma^{(2)}+2A^{(0)}(\gamma^{(1)})^2\partial_{\Delta}\big]\,\mathcal{B}_{n,\ell}(x,\xb)\\
						 +&\log^0(U)\sum \big[A^{(2)}+2A^{(1)}\gamma^{(1)}\partial_{\Delta}+2A^{(0)}\gamma^{(2)}\partial_{\Delta}+2A^{(0)}(\gamma^{(1)})^2\partial_{\Delta}^2\big]\,\mathcal{B}_{n,\ell}(x,\xb)\,,
\end{split}
\end{align}
where for brevity we have suppressed all summation indices. The above sums run over even twists $\tau^{(0)}=2n\geq4$\,\footnote{
	This statement is strictly speaking only true for the logarithmic terms (in $U$) in equation \eqref{eq:log(u)_stratification}, which indeed receive contributions only from unprotected double- and higher-trace operators. In the analytic parts, i.e. the coefficients of $\log^0(U)$, the long part of the free theory correlator gives some twist 2 contributions, and due to multiplet recombination in the twist-2 sector there is an ambiguity in these OPE coefficients, which however does not modify higher-twist contributions. We refer the reader to the discussion in \cite{Alday:2021ajh} for more details.
	}
and even/odd spins $\ell$, depending on the parity of the channel $\mathbf{a}$. Note that due to the degeneracy in the spectrum of double-trace operators \eqref{eq:double-trace_ops}, the above expressions for the OPE data have to be understood as ``averaged'' quantities. A detailed explanation on how to resolve this operator mixing will be given in Section \ref{sec:leading_log}.

Lastly, we recall that the difference between the long part $\mathcal{K}$ and the reduced correlator $\mathcal{H}$ is simply a part of the free theory correlator $\mathcal{G}_0$, which contributes only up to order $1/N$ and furthermore contains no $\log(U)$ contributions. The terms of order $1/N^2$ or higher in the expansion \eqref{eq:large_N_exp_K} are thus equal to the corresponding large $N$ expansion of the reduced correlator:
\begin{align}
	\mathcal{H}_\mathbf{a}^{(m)}(x,\xb) = \mathcal{K}_\mathbf{a}^{(m)}(x,\xb)\,,\quad m\geq2\,.
\end{align}
This is equivalent to saying that no states of twist $\tau<4$ contribute to the superconformal block decomposition starting from the one-loop correction $\mathcal{H}^{(2)}$. Moreover, the same holds for the $\log(U)$ part at tree-level, i.e. we have
\begin{align}
	\mathcal{H}_\mathbf{a}^{(1)}(x,\xb)\vert_{\log(U)} = \mathcal{K}_\mathbf{a}^{(1)}(x,\xb)\vert_{\log(U)}\,.
\end{align}
In the following, we will often use this equivalence and refer to the block decomposition of the reduced correlator $\mathcal{H}$ instead of the long part $\mathcal{K}$.

\subsection{Review of tree-level correlators}\label{sec:tree_level_review}
At this point, let us give a short review of what is currently known about the tree-level Veneziano amplitude in AdS$_5\times$S$^3$. Much like its flat space counterpart, it is best represented in terms of colour-ordered amplitudes
\begin{align}\label{eq:M1_colour_ordered}
	\mathcal{M}^{(1)} =  \mathcal{M}^{(1)}(1234) \Tr(1234) + \mathcal{M}^{(1)}(1243) \Tr(1243)+ \mathcal{M}^{(1)}(1324) \Tr(1324)\,,
\end{align}
where we used the short-hand notation
\begin{align}
	\Tr(1234)\equiv\Tr(T^{I_1}T^{I_2}T^{I_3}T^{I_4})\,,
\end{align}
and similarly for the other permutations. In the above, $T^I$ denotes the generators of $SO(8)$ in the fundamental representation, which we normalise as $\Tr(T^{I}T^{J}) = \delta^{IJ}$. Note that due to their antisymmetry, there are only three independent colour traces. Their decomposition into the irreps $\mathbf{a}$ given in \eqref{eq:adjxadj} reads
\begin{align}\label{eq:tr_decomp}
\begin{split}
	\Tr(1234)&=\big\{\tfrac{7}{2},\tfrac{3}{2},0,0,0,\tfrac{3}{2},0\big\},\\
	\Tr(1243)&=\big\{\tfrac{7}{2},\tfrac{3}{2},0,0,0,-\tfrac{3}{2},0\big\},\\
	\Tr(1324)&=\big\{\tfrac{1}{2},\tfrac{1}{2},-1,-1,\tfrac{1}{2},0,0\big\}.
\end{split}
\end{align}

Recall that at large $\lambda$ the tree-level Mellin amplitude $\mathcal{M}^{(1)}$ admits an expansion in $1/\lambda$ of the form
\begin{align}\label{eq:M1_lambda_exp_2}
	\mathcal{M}^{(1)} = \mathcal{M}^{(1,0)} + \lambda^{-1}\mathcal{M}^{(1,2)} + \lambda^{-\frac{3}{2}}\mathcal{M}^{(1,3)} + \lambda^{-2}\mathcal{M}^{(1,4)} + O(\lambda^{-\frac{5}{2}})\,,
\end{align}
and each colour-ordered amplitude from \eqref{eq:M1_colour_ordered} inherits an analogous expansion. Here we will restrict ourselves to the correlator of lowest dimensions, $p_i=2$. Correlators with arbitrary external charges have been constructed in \cite{Glew:2023wik}, whose results we summarise in Appendix \ref{app:corrgencharges}.
The first term is the field-theory contribution, computed in \cite{Alday:2021odx}:\footnote{It is sufficient to only quote the result for the colour-ordered amplitude $\mathcal{M}^{(1)}(1234)$, as the others are easily obtained by crossing. Explicitly, one has
\begin{align}
	\mathcal{M}^{(1)}(1243) = \mathcal{M}^{(1)}(1234)\vert_{t\leftrightarrow u}\,,\quad \mathcal{M}^{(1)}(1324) = \mathcal{M}^{(1)}(1234)\vert_{s\leftrightarrow u}\,.
\end{align}}
\begin{align}
	\mathcal{M}^{(1,0)}(1234) =-\frac{2}{(s+1)(t+1)}\,.
\end{align}

Following the field theory term, there is an infinite tower of higher-derivative corrections weighted by half-integer powers of $1/\lambda$, which have recently been considered in \cite{Glew:2023wik,Behan:2023fqq}. The first non-vanishing correction stems from an $F^4$ contact term at order $\lambda^{-1}$, whose Mellin amplitude is just a constant,
\begin{align}
	\mathcal{M}^{(1,2)}(1234) = 192\zeta_2\,.
\end{align}
The next term, the $D^2F^4$ correction at order $\lambda^{-3/2}$, is linear in the Mellin variables and reads
\begin{align}\label{eq:M13}
	\mathcal{M}^{(1,3)}(1234) = -3072\zeta_3\,(u+1)\,.
\end{align}
Let us also comment on a previously unnoticed property of this term: just like the field-theory amplitude $\mathcal{M}^{(1,0)}$, the $\lambda^{-3/2}$ correction satisfies the $U(1)$ decoupling identity and the BCJ relations, see \cite{Drummond:2022dxd} for a description of these relations for the field-theory contribution. As a consequence, instead of using the single-trace basis \eqref{eq:M1_colour_ordered}, one can rewrite $\mathcal{M}^{(1,3)}$ as
\begin{align}\label{eq:M13_c}
	\mathcal{M}^{(1,3)}(s,t) = 512\zeta_3\big[(t-u)c_s+(u-s)c_t+(s-t)c_u\big] \equiv \mathcal{M}_s\,c_s + \mathcal{M}_t\,c_t + \mathcal{M}_u\,c_u\,,
\end{align}
where the colour factors $c_{s,t,u}$ are given by products of structure constants: $c_s=f^{I_1I_2K}f^{KI_3I_4}$, $c_t=f^{I_1I_4K}f^{KI_2I_3}$ and $c_u=f^{I_1I_3K}f^{KI_4I_2}$.\footnote{In terms of the trace-basis \eqref{eq:tr_decomp}, these colour structures read
\begin{align}
	\tfrac{1}{2}c_s=\Tr(1234)-\Tr(1243)\,,~~ \tfrac{1}{2}c_t=\Tr(1324)-\Tr(1234)\,,~~ \tfrac{1}{2}c_u=\Tr(1243)-\Tr(1324)\,,
\end{align}
but note that due to the Jacobi identity $c_s+c_t+c_u=0$ these relations can not be inverted.}
Note that this rewriting crucially depends on the precise value of the constant term in \eqref{eq:M13}, as fixed from supersymmetric localisation in \cite{Behan:2023fqq}. Rewritten in this way, one finds that the so-called colour-kinamtics duality between the kinematical factors $\mathcal{M}_{s,t,u}$ and the colour structures $c_{s,t,u}$ holds, which (even in flat space) is not true for a generic term in the low energy expansion of the tree-level string amplitude. As we will see later on, this special property of the tree-level term has implications for the one-loop $\lambda^{-3/2}$ correction, which turns out to have a unique colour structure among the other one-loop terms.

Finally, the $D^4F^4$ correction at order $\lambda^{-2}$ is currently known up to four undetermined parameters. Specialising the result from \cite{Glew:2023wik} to $p_i=2$, we have
\begin{align}\label{eq:M14}
\begin{split}
	\mathcal{M}^{(1,4)}(1234) &= \frac{256\pi^4}{15}\big(5(7s^2+7t^2+u^2)+11(7s+7t+u)+90\big)\\
							  &\quad+3072a_2-12288b_1u-12288e_1(u+\tfrac{6}{7})+6144f_1(2u+\tfrac{15}{7})\,,
\end{split}
\end{align}
where $a_2$, $b_1$, $e_1$ and $f_1$ are the four free parameters. Note that the above expression for the $\langle2222\rangle$ correlator really contains only \emph{two independent} parameters; a linear term in $u$ and a constant. The four parameters are only independent in the correlator of arbitrary external charges. We have explicitly kept all four parameters because in the computation of the one-loop leading log -- as explained in great detail in the next section -- we will in fact need knowledge about the family of $\langle22pp\rangle$ correlators. As such, we will find that all \textit{four} parameters will propagate into leading log at order $\lambda^{-2}$.

\section{Predicting the leading log: from trees to loops}\label{sec:leading_log}
In this section we recall the relation between tree-level and one-loop discontinuities dictated by the OPE. In particular, we will show that, at the first few orders, it is possible to relate tree-level and one-loop correlators via a fourth order differential operator. This is ultimately due to the fact that the double-trace spectrum emerging from tree-level correlators inherits from the amplitude an 8-dimensional structure. Before explaining this in detail, it is necessary recall the relation between OPE coefficients and data of two-particle operators.

The OPE analysis will be carried in the basis of irreps and only after having constructed the amplitude, we will switch to a suitable colour basis.  To avoid cluttering the notation, we will often drop the subscript specifying the different $SO(8)$ representations, when this does not create confusion.

\subsection{OPE equations and double-trace data}\label{sec:OPE_eqs}
As mentioned already, at large $N$, many double-trace operators of the schematic form
\begin{equation}
	\mathcal{O}_p \square^{\tfrac{\tau-p-q}{2}} \partial_\ell \mathcal{O}_q
\end{equation}
are degenerate, where in the above expression we have dropped the flavour indices for simplicity. The number of degenerate operators is equal to the number of points filling a certain rectangle \cite{Drummond:2022dxd}, which we recall in Appendix \ref{sec:treeleveunmix}. In the singlet representation of $SU(2)_R\times SU(2)_L$ there are $n-1$ of these, as listed in equation \eqref{eq:double-trace_ops}.
Double-trace operators generically mix when interactions are turned on, and for this reason the OPE equations are better organised into matrix equations. The purpose of this subsection is to rewrite  \eqref{eq:log(u)_stratification} in this fashion by taking into account the mixing.

Let us denote by $\mathcal{S}_{p}$ the set of true scaling eigenstate, with $p=2,\ldots n$. We recall that the scaling dimension admits the expansion
\begin{equation}
	\tau_{\mathcal{S}}= 2 n+ \frac{2}{N} \gamma_{\mathcal{S}}^{(1)}+ O \left(\frac{1}{N^2}\right).
\end{equation} 
Similarly, we can expand the three-point couplings $C_{pp\mathcal{S}_q}=\langle \mathcal{O}_p  \mathcal{O}_p \mathcal{S}_q \rangle$ of the two-particle operators with two operators $\mathcal{O}_p$ in $\frac{1}{N}$:
\begin{equation}
	C_{pp\mathcal{S}_q}=C_{pp\mathcal{S}_q}^{(0)}+ \frac{1}{N} C_{pp\mathcal{S}_q}^{(1)}+ O \left(\frac{1}{N^2}\right).
\end{equation}

The mixing in the singlet can be solved by considering the $\langle \mathcal{O}_p \mathcal{O}_p \mathcal{O}_q\mathcal{O}_q \rangle$ family of correlators with $p,q=2,\ldots n$, and arrange the various correlators in a $(n-1) \times (n-1)$ matrix. The latter admits the following block expansion:
\begin{align}
\begin{split}
	\mathbf{\mathcal{H}}_{ppqq}^{(0)} &=\log^0(U)\sum_{n,\ell} \mathbf{L}_{n,\ell}^{(0)} \,\mathcal{B}_{n,\ell}(x,\xb)\,,\\[3pt]
		\mathbf{\mathcal{H}}_{ppqq}^{(1)} &=\log^1(U)  \sum_{n,\ell}  \mathbf{M}_{n,\ell}^{(1)} \, \mathcal{B}_{n,\ell}(x,\xb) + \cdots, \\[3pt]
	\mathbf{\mathcal{H}}_{ppqq}^{(2)} &= \log^2(U) \sum_{n,\ell} \mathbf{N}_{n,\ell}^{(2)}\, \mathcal{B}_{n,\ell}(x,\xb)+\cdots,
\end{split}
\end{align}
where the dots are terms containing lower powers of $\log(U)$ which are not relevant for us, and  $\mathbf{L}^{(0)},\mathbf{M}^{(1)},\mathbf{N}^{(2)}$ are matrices of CPW coefficients of the leading $\log(U)$ projection of the correlator at each order in $1/N$.
Consistency with the OPE leads to the following set of equations:
\begin{align}\label{OPEequ}
	\mathbf{L}^{(0)} &= \mathbf{C}^{(0)} {\mathbf{C}^{(0)}}^T ,\\
	\mathbf{M}^{(1)} &= \mathbf{C}^{(0)} \gamma^{(1)} {\mathbf{C}^{(0)}}^T ,\\
	\mathbf{N}^{(2)} &= \frac{1}{2} \mathbf{C}^{(0)} (\gamma^{(1)})^2 {\mathbf{C}^{(0)}}^T.
\end{align}
Here, $\mathbf{C}^{(0)}$ is a $(n-1) \times (n-1)$ matrix of three-point functions $C_{pp \mathcal{S}_q}$ with $p,q=2,\ldots n$, and $\gamma^{(1)}$ is a $(n-1) \times (n-1)$ diagonal matrix of anomalous dimensions with elements $\gamma^{(1)}_p$, $p=2,\ldots,n$.

Let us now expand three-point functions and anomalous dimensions in $\frac{1}{\sqrt{\lambda}}$
\begin{align}
	C_{pp\mathcal{S}}^{(0)} &=C_{pp\mathcal{S}}^{(0,0)}+C_{pp\mathcal{S}}^{(0,2)}\lambda^{-1} +C_{pp\mathcal{S}}^{(0,3)}\lambda^{-\frac{3}{2}}+C_{pp\mathcal{S}}^{(0,4)} \lambda^{-2}+ O( \lambda^{-\frac{5}{2}})\,,\\
	\gamma_{p}^{(1)} &= \gamma_{p}^{(1,0)}+ \gamma_{p}^{(1,2)}\lambda^{-1} + \gamma_{p}^{(1,3)}\lambda^{-\frac{3}{2}}+ \gamma_{p}^{(1,4)} \lambda^{-2}+ O( \lambda^{-\frac{5}{2}})\,,
\end{align}
and plug the expansion into the OPE equations \eqref{OPEequ}.
In the field theory limit, the relevant unmixing equations are
\begin{align}\label{OPEtreelev}
	\mathbf{L}^{(0)} &= \mathbf{C}^{(0,0)} {\mathbf{C}^{(0,0)}}^T,\\
	\mathbf{M}^{(1,0)} &= \mathbf{C}^{(0,0)} \gamma^{(1,0)} {\mathbf{C}^{(0,0)}}^T,
\end{align}
which return three-point functions and anomalous dimension of the unmixed eigenstates $\mathcal{S}_p$. This eigenvalue problem has been solved in \cite{Drummond:2022dxd} and we review the results for the singlet in next section, as well as the general unmixing for all channels in Appendix \ref{sec:treeleveunmix}. In the appendix we also write down explicitly the 
analogous tree-level equations for the first few orders in $\frac{1}{\sqrt{\lambda}}$.

In this paper we are interested in the ($\log^2(U)$ projection of the) $\frac{1}{N^2}$ contribution, which is fully fixed by tree-level data.
In particular, the first two orders read:
\begin{align}
	O(\lambda^{-1}):\qquad\mathbf{N}^{(2,2)} &= \mathbf{C}^{(0,0)} \gamma^{(1,0)}\gamma^{(1,2)} {\mathbf{C}^{(0)}}^T+\frac{1}{2}\left( \mathbf{C}^{(0,0)} \gamma^{(1,0)} \gamma^{(1,0)} {\mathbf{C}^{(0,2)}}^T +\text{tr} \right),\\
	O(\lambda^{-\frac{3}{2}}):\qquad \mathbf{N}^{(2,3)} &= \mathbf{C}^{(0,0)}  \gamma^{(1,0)}\gamma^{(1,3)} {\mathbf{C}^{(0)}}^T+\frac{1}{2}\left( \mathbf{C}^{(0,0)} \gamma^{(1,0)} \gamma^{(1,0)} {\mathbf{C}^{(0,3)}}^T +\text{tr} \right).
\end{align}

At order $\lambda^{-2}$ there is one more contribution one needs to take into account, which comes from squaring the anomalous dimensions at $\lambda^{-1}$. In sum:
\begin{align}\label{eq:N24_OPE}
\mathbf{N}^{(2,4)} &= \mathbf{C}^{(0,0)} \gamma^{(1,0)} \gamma^{(1,4)} {\mathbf{C}^{(0,0)}}^T + \frac{1}{2}\left( \mathbf{C}^{(0,0)} \gamma^{(1,0)} \gamma^{(1,0)} {\mathbf{C}^{(0,4)}}^T +\text{tr} \right)+ \frac{1}{2} \left( \mathbf{C}^{(0,0)} {\gamma^{(1,2)}} {\gamma^{(1,2)}} {\mathbf{C}^{(0,0)}}^T \right) \notag \\
& =\frac{1}{2}\left( \mathbf{M}^{(1,4)} \, {\mathbf{L}^{(0)}}^{-1} \mathbf{M}^{(1,0)} + \text{tr}  \right)+  \frac{1}{2}\left( \mathbf{M}^{(1,2)} \, {\mathbf{L}^{(0)}}^{-1} \mathbf{M}^{(1,2)}   \right),  
\end{align}
where in the second equality we have rewritten it in terms of tree-level CPW coefficients.
Note that this latter rewriting is quite useful as it allows to avoid the computations of three-point functions and anomalous dimensions.

It is clear from the unmixing equations that the computation of $\mathbf{N}^{(2,m)}$ and thus of the one-loop correlators, relies on the knowledge of tree-level CPW coefficients.
However, in this particular theory it is possible to avoid a direct computation of the latter, at least for the first few orders, essentially because tree-level correlators enjoy a hidden 8-dimensional conformal symmetry, which has precise consequences on the spectrum of double-trace operators. Explaining this is the purpose of the remaining part of the section.

\subsection{8d hidden conformal symmetry in the double-trace spectrum}
The field-theory correlator possesses an hidden 8-dimensional conformal symmetry \cite{Alday:2021odx}. At the level of the amplitude, this means that the correlator for arbitrary charges can be obtained from $\mathcal{H}_{2222}$ by promoting 4-dimensional to 8-dimensional distances. When string corrections are added, the symmetry is broken. However, the associated amplitudes are still governed by an 8-dimensional principle \cite{Glew:2023wik}.
We will not go into the details of the consequences of the hidden symmetry on the correlators as they are not relevant to this paper which will mainly focus on the correlator with minimal charges. 

As we mentioned before, at order $1/N$, i.e. tree-level, all degenerate double-trace operators receive different corrections to their dimension and the degeneracy is generically lifted.
In theories with hidden symmetries, such as the one we are interested in, these anomalous dimensions can be explicitly written in terms of rational functions.\footnote{This was originally noticed in \cite{Aprile:2017xsp,Aprile:2018efk} for AdS$_5\times$S$^5$ and later found to be true in all other known theories with a higher-dimensional structure \cite{Aprile:2021mvq,Abl:2021mxo,Drummond:2022dxd}.} We should perhaps remark that this is highly non-trivial as, by definition, the anomalous dimensions are the zeros of an $m-$th order characteristic polynomial, where the order depends on $\tau$ and the R-symmetry representation exchanged, and in general cannot be written in terms of radicals.
Shortly, the reason why this remarkable simplification occurs is that from an higher-dimensional perspective there is really one operator per each twist and thus no mixing \cite{Caron-Huot:2018kta}.
The explicit form of the anomalous dimensions in AdS$_5\times$S$^3$ has been worked out in \cite{Drummond:2022dxd}. Taking into account the $SO(8)$ flavour decomposition, we have, for the $SU(2)_L \times SU(2)_R$ singlet representation,
\begin{equation}\label{eq:gamma10_singlet}
\gamma_{\mathbf{a},p}^{(1,0)}= 
\begin{pmatrix}
-6\\ -2\\ -2\\ -2\\ 1\\\hline -3 \\ 0
\end{pmatrix}
\frac{\delta_{\tau,\ell}^{(4)}}{(\ell_{8d}+1)_4}\equiv  v_{\mathbf{a}}\, \frac{\delta_{\tau,\ell}^{(4)}}{(\ell_{8d}+1)_4}\,,
\end{equation}
where the \emph{effective 8-dimensional spin} reads 
\begin{equation}\label{8dspin}
\ell_{8d}= \ell +2(p-2) \,,
\end{equation}
and $\delta_{\tau,\ell}^{(4)}$ is the eigenvalue of $\Delta^{(4)}$ which we introduced earlier in equation \eqref{eq:eigenvalue_d4}.
Here, we have also defined the colour vector $v_{\mathbf{a}}$ for later convenience. The name 8-dimensional spin is justified by the fact that this quantity behaves effectively as a spin in 8d flat space. In fact, by adapting the argument of \cite{Caron-Huot:2018kta}, one can explicitly check that the denominator $(\ell+1)_4$ is exactly the same quantity which appears in the partial wave expansion of the 8-dimensional gluon amplitude. We will see this explicitly in Section \ref{sec:flat-space-limit}.

Let us now consider string corrections. We will argue that $\ell_{8d}$ in fact retains its meaning as a 8d spin in flat space \emph{at all orders in} $\lambda^{-\frac{1}{2}}$, and not just in the field-theory limit. Now, since the spin in the flat-space Veneziano amplitude is bounded from above, i.e. at order $\alpha'^{m+2}$ one has $\ell \leq m-2-\frac{1\pm(-1)^{m+1}}{2}$ (see Section \ref{sec:flat-space-limit} for more details), we expect the analogous property to hold also in AdS$_5\times$S$^3$: at order $\lambda^{-\frac{m}{2}}$, we expect
\begin{align}\label{inequalL}
	\ell_{8d} \leq m-2-\frac{1\pm(-1)^{m+1}}{2}\,,
 \end{align}
where the $+$ ($-$) sign refers to symmetric (antisymmetric) $SO(8)$ irreps. The idea is that the inequality \eqref{inequalL} dictates which double-trace operators receive a non-zero correction. In other words, three-point functions and anomalous dimensions of operators whose quantum numbers are outside of the above bound, do vanish:
\begin{equation}\label{eq:conject8d}
\begin{cases}
&\,\gamma_{\mathbf{a},p}^{(1,m)}=0\,,    \\
&C_{qq\mathcal{S}_p}^{(0,m)}=0\,,	 
\end{cases}
 \qquad \text{ for} \quad \ell_{8d} > m-2-\frac{1\pm(-1)^{m+1}}{2}\,.
\end{equation}

An analogous condition was originally put forward in AdS$_5\times$S$^5$   background \cite{Drummond:2019odu,Drummond:2020dwr} and refined in \cite{Aprile:2020mus}, where it was found that (a 10d version of) the above inequality gives rise to tree-level amplitudes which are in perfect agreement with all results available in literature, and in particular with the effective field theory approach of \cite{Abl:2020dbx}.
Since the argument is based on the existence of a higher-dimensional symmetry, it is natural to expect that a similar story should hold in this background too.
With the tree-level amplitudes at our disposal, we have explicitly tested the predictions \eqref{eq:conject8d} to order $\lambda^{-2}$, see Appendix \ref{sec:treeleveunmix} for details on the computation.

In the following, we will focus on the consequences of the bound and show how it simplifies the computation of the one-loop leading log. Let us illustrate this explicitly for the first few cases.
\paragraph{Order $\lambda^{-1}/N$:} The bound \eqref{inequalL} reads $\ell_{8d}=0$. This in turn implies that there is only one operator turned on, i.e.  the one with $p=2$ and $\ell=0$. Moreover, the corrected three-point function at this order vanish, ${\mathbf{C}^{(0,2)}}=0$.
\paragraph{Order $\lambda^{-3/2}/N$:} At this order this situation is similar, the main difference being that here we have both symmetric and antisymmetric amplitudes.
In both cases the bound \eqref{inequalL} predicts again one anomalous dimension turned on, with $p=2$ and $\ell=0$ ($\ell=1$) in the symmetric (antisymmetric) amplitude. Also here, the bound predicts ${\mathbf{C}^{(0,3)}}=0$, for all values of $\ell$.

\paragraph{Order $\lambda^{-2}/N$:} The antisymmetric amplitude again gives rise to a spectrum with only one operator getting a non-zero correction to its anomalous dimension, i.e. the one with $p=2, \ell=1$;  ${\mathbf{C}^{(0,4)}}\vert_{\ell=1}=0$.
On the other hand, the symmetric amplitudes will now give rise to a spectrum with three non-zero anomalous turned on. This is immediate from \eqref{inequalL}; in particular the non-zero anomalous dimensions that satisfy the bound have labels $(p=3,\ell=0)$, $(p=2,\ell=2)$ with $\ell_{8d}=2$, and  $(p=2,\ell=0)$ with $\ell_{8d}=0$. Moreover, the corrected three-point functions for $\ell=0$ are non vanishing at this order, ${\mathbf{C}^{(0,4)}}\vert_{\ell=0}\neq 0$.

We will not write down the explicit form of the anomalous dimensions as it will not be needed for the computation of one-loop string amplitudes. 
For sake of completeness, let us just mention that the existence of a hidden symmetry -- and its breaking due to string corrections -- reflects on the spectrum of CFT data. This is not visible in the singlet $SU(2)_L\times SU(2)_R$ representation but one needs to consider non-trivial reps with higher $SU(2)_L\times SU(2)_R$ spins. In a nutshell, the field-theory anomalous dimensions do not completely resolve the free-theory mixing but possess a residual degeneracy, which is a consequence of the associated tree-level correlator exhibiting a hidden symmetry \cite{Caron-Huot:2018kta}. 
When adding string corrections, the hidden symmetry is broken and the accidental degeneracy is sequentially resolved, as shown in \cite{Aprile:2020mus} for the analogous AdS$_5\times$S$^5$ case.
In Appendix \ref{sec:treeleveunmix} we show the mechanism at work for the first few orders. The interested reader can find the explicit form of these $\lambda$-corrected anomalous dimensions in \cite{Santagata:2022hga}.

\subsection{One-loop leading logs from tree-level}\label{sec:leading_log_one-loop}
We end this section by showing how the observations we made on the spectrum of double-trace operators allow to greatly simplify the computation of the leading logs at one-loop, at least for the first few cases.
In fact, we are going to explain how the one-loop leading log is related to the tree-level discontinuity via application of the operator $\dfour$.

\paragraph{Order $\lambda^{-1}/N^2$:} As discussed above, only the operator with $p=2$ and $\ell=0$ acquires an anomalous dimension at this order, and the three-point functions ${\mathbf{C}^{(0,2)}}$ vanish. We therefore have
\begin{align}\label{eq:llog_2_sum}
\begin{split}
	\mathcal{H}_\mathbf{a}^{(2,2)}\big|_{\log^2 (U)} &= \sum_{n,\ell}\Big[ \sum_p  C_{22 \mathcal{S}_p}^{(0,0)} \gamma_{\mathbf{a},p}^{(1,0)}\gamma_{\mathbf{a},p}^{(1,2)}  C_{22 \mathcal{S}_p}^{(0,0)}\Big]_{n,\ell}\, \mathcal{B}_{n,\ell}(x,\xb)\\
	&=\sum_{n} \Big[C_{22 \mathcal{S}_2}^{(0,0)} \gamma_{\mathbf{a},2}^{(1,0)}\gamma_{\mathbf{a},2}^{(1,2)} C_{22 \mathcal{S}_2}^{(0,0)}\Big]_{\ell=0}\, \mathcal{B}_{n,0}(x,\xb)\,,
\end{split}
\end{align}
where in the second equality we have used the fact that $\gamma_{\mathbf{a},p}^{(1,2)}\propto \delta_{p,2}\,\delta_{\ell,0}$.
Now, from equation \eqref{eq:gamma10_singlet} we have that $\gamma_{\mathbf{a},p=2}^{(1,0)}\vert_{\ell=0} = \frac{1}{24}v_\mathbf{a}\, \delta_{n,0}^{(4)}$, such that
\begin{align}
\begin{split}
	\mathcal{H}_{\mathbf{a}}^{(2,2)}\big|_{\log^2 (U)}& =  \frac{1}{24} v_{\mathbf{a}} \sum_{n} \Big[C_{22 \mathcal{S}_2}^{(0,0)} \delta_{n,0}^{(4)}\gamma_{\mathbf{a},2}^{(1,2)} C_{22,2}^{(0,0)}\Big]_{\ell=0} \, \mathcal{B}_{n,0}(x,\xb) \\
	&=  \frac{1}{24}\,v_{\mathbf{a}}\, \dfour \Big( \sum_{n} C_{22 \mathcal{S}_2}^{(0,0}\gamma_{\mathbf{a},2}^{(1,2)} C_{22 \mathcal{S}_2}^{(0,0)} \, \mathcal{B}_{n,0}(x,\xb) \Big),
\end{split}
\end{align}
where in the second step we used the eigenvalue equation \eqref{eq:eigenvalue_blocks} to trade a power of $\delta_{n,\ell}^{(4)}$ for the differential operator $\dfour$. In the remaining term in brackets, we recognise the partial wave decomposition of the $\log(U)$-part of the tree-level amplitude $\mathcal{H}_\mathbf{a}^{(1,2)}$. We thus find
\begin{align}\label{eq:llog_2}
	\mathcal{H}_{\mathbf{a}}^{(2,2)}\big|_{\log^2 (U)} =  \frac{1}{24} \, v_{\mathbf{a}} \, \dfour \mathcal{H}_{\mathbf{a}}^{(1,2)}\big|_{\log (U)}\,,
\end{align}
where we recall the definition of the colour vector $v_\mathbf{a}=\{-6,-2,-2,-2,1,-3,0\}^{\text{T}}$ from equation \eqref{eq:gamma10_singlet}. The above is our desired result: the one-loop leading log is simply obtained by application of $\dfour$ on the tree-level result!

\paragraph{Order $\lambda^{-3/2}/N^2$:}
The situation is very similar at this order. As before, there is a single 8-dimensional spin which is exchanged, namely the operator with $p=2$ and $\ell_{8d}=0$ ($\ell_{8d}=1$) for the symmetric (antisymmetric) irreps. Consequently, the one-loop leading log is computed by
\begin{align}\label{eq:llog_3}
	\mathcal{H}_{\mathbf{a}}^{(2,3)}\big|_{\log^2 ( U)} =
	\begin{cases}
	\frac{1}{24}v_\mathbf{a} \, \dfour \, \mathcal{H}_{\mathbf{a}}^{(1,3)}\vert_{\log (U)}\,, &\text{ for $\mathbf{a}$ symmetric}\,,\\[5pt]
	\frac{1}{120}v_\mathbf{a} \, \dfour \, \mathcal{H}_{\mathbf{a}}^{(1,3)}\big|_{\log (U)}\,, &\text{ for $\mathbf{a}$ antisymmetric}\,,
	\end{cases}
\end{align}
where for the antisymmetric irreps we have used the fact that
$\gamma_{2,\mathbf{a}}^{(1,0)}\vert_{\ell=1} = \frac{1}{120}v_\mathbf{a}\, \delta_{n,1}^{(4)}$\,.

\paragraph{Order $\lambda^{-2}/N^2$:} As we have seen from the OPE prediction in equation \eqref{eq:N24_OPE}, there are two distinct contributions at this order. We find it useful to consider them separately, and write the full leading log as a sum of two parts:
\begin{align}
	\mathcal{H}_{\mathbf{a}}^{(2,4)}\vert_{\log^2( U)} \equiv H_{\mathbf{a}}' + H_{\mathbf{a}}''\,.
\end{align}

The first part, coming from the order $\lambda^{-1}$ anomalous dimension squared, contains again only spin $\ell_{8d}=0$ exchanges. Moreover, we can make us of the fact that this anomalous dimension can be written as the eigenvalue $\delta_{n,\ell=0}$ squared, see equation \eqref{eq:gamma12} in the appendix for more details. This allows the first part of the leading log to be written concisely as
\begin{align}
	H'_\mathbf{a} = -\frac{1}{2}\cdot\frac{2\zeta_2}{15} \, \tilde{v}_\mathbf{a} \, \big(\dfour\big)^2 \mathcal{H}_{\mathbf{a}}^{(1,2)}\big|_{\log (U)}\,,
\end{align}
where the colour vector $\tilde{v}_\mathbf{a}$ is given by $\tilde{v}_\mathbf{a}=\{15,7,-2,-2,1,0,0\}^{\text{T}}$.

The second part, $H''_\mathbf{a}$, requires some more care. In the antisymmetric irreps, we still only have a single 8d spin exchanged, $\ell_{8d}=1$. As in \eqref{eq:llog_3}, we thus have
\begin{equation}
	H''_\mathbf{a} =  \frac{1}{120} v_\mathbf{a} \, \dfour \, \mathcal{H}_{\mathbf{a}}^{(1,4)}\vert_{\log (U)}\,,  \qquad  \text{ for $\mathbf{a}$ antisymmetric}\,.
\end{equation}
Symmetric irreps contain exchanges of states with more than a single 8d spin however: as dictated by \eqref{inequalL}, we find contributions from $\ell_{8d}=0,2$. This prevents us from computing the one-loop leading log from the tree-level correlator via $\dfour$, since contributions from different 8d spins are weighted by different factors.\footnote{Splitting OPE decomposition of the tree-level $\log(U)$ part into contributions from different 8d spins, we have
\begin{equation}
	\mathcal{H}_{\mathbf{a}}^{(1,4)}\vert_{\log (U)} = \sum_{\substack{n,\ell,p \\ \ell+2p=4}}M_{n,\ell}^{(1,4)} \mathcal{B}_{n,\ell}(x,\xb)+\sum_{\substack{n,\ell,p \\ \ell+2p=6}}M_{n,\ell}^{(1,4)} \mathcal{B}_{n,\ell}(x,\xb)\,.
\end{equation}
To obtain the one-loop leading log from this, we would like to insert the field-theory anomalous dimension $\gamma^{(1,0)}$ by acting with $\dfour$ on this expression. However, due to the denominator $(\ell_{8d}+1)_4$ in \eqref{eq:gamma10_singlet}, the two terms will differ by a factor and will therefore not reassemble into $H''_\mathbf{a}$.
}
Instead, we need to directly resum the OPE prediction, which for this term reads
\begin{align}\label{eq:H''_resummation}
	H''_\mathbf{a} = \frac{1}{2}\sum_{n,\ell} \left( \mathbf{M}^{(1,4)} \,{\mathbf{L}^{(0)}}^{-1} \mathbf{M}^{(1,0)} + \text{tr}  \right) \mathcal{B}_{n,\ell}(x,\xb)\,.
\end{align}
As explained in Section \ref{sec:OPE_eqs}, for this we need to consider \textit{matrices} of CPW coefficients, which collect the OPE data of correlators of the type $\langle ppqq \rangle$. Since we are interested in the $\langle 2222 \rangle$ correlator only (the upper-left corner of these matrices), we actually only need to consider the $\langle22pp\rangle$ family of correlators.

We have explicitly performed the OPE sum \eqref{eq:H''_resummation} for the symmetric irreps. We omit reproducing the explicit result here, as it can be easily obtained from e.g. the position space results for the one-loop correlator discussed in Section \ref{sec:oneloopamp}. It is important to note that the leading log prediction $H''_{\mathbf{a}}$ inherits the four free parameters from the tree-level amplitude $\mathcal{M}^{(1,4)}$, c.f. \eqref{eq:M14}.

\subsection{The colour structure of loop-amplitudes}\label{sec:colour_structure_loops}
So far, in the above derivation of the one-loop leading logs, we have worked in the basis of $SO(8)$ irreps \eqref{eq:vector_notation}. In fact, this is the natural basis which is compatible with the OPE decomposition, on which our construction of the leading logs relies.

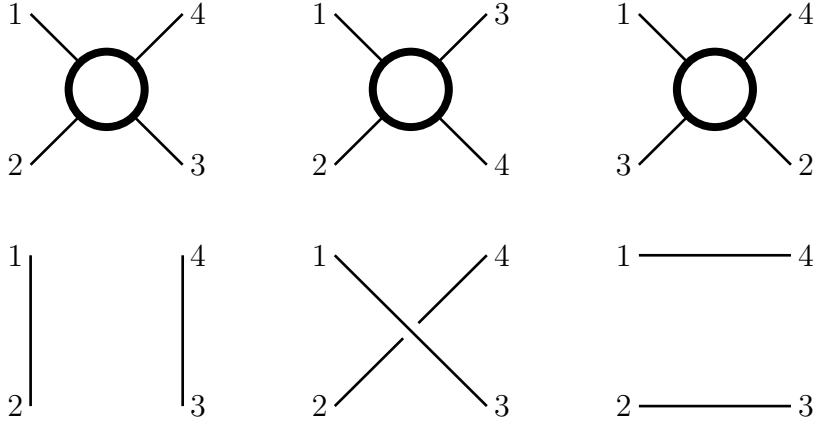
\begin{figure}
\centering
\begin{tikzpicture}[scale=1.0]
\coordinate (a) at (0,3.2);
\coordinate (b) at (4,3.2);
\coordinate (c) at (8,3.2);
\coordinate (d) at (0,0);
\coordinate (e) at (4,0);
\coordinate (f) at (8,0);
\draw[line width=3pt] ($(a)+(1,1)$) circle (0.5);
\draw[line width=1pt] (a) -- ($(a)+(0.65,0.65)$);
\draw[line width=1pt] ($(a)+(2,0)$) -- ($(a)+(2-0.65,0.65)$);
\draw[line width=1pt] ($(a)+(0,2)$) -- ($(a)+(0.65,2-0.65)$);
\draw[line width=1pt] ($(a)+(2,2)$) -- ($(a)+(2-0.65,2-0.65)$);
\node at ($(a)+(0,2)-(0.2,0)$) {1};
\node at ($(a)-(0.2,0)$) {2};
\node at ($(a)+(2,0)+(0.2,0)$) {3};
\node at ($(a)+(2,2)+(0.2,0)$) {4};
\draw[line width=3pt] ($(b)+(1,1)$) circle (0.5);
\draw[line width=1pt] (b) -- ($(b)+(0.65,0.65)$);
\draw[line width=1pt] ($(b)+(2,0)$) -- ($(b)+(2-0.65,0.65)$);
\draw[line width=1pt] ($(b)+(0,2)$) -- ($(b)+(0.65,2-0.65)$);
\draw[line width=1pt] ($(b)+(2,2)$) -- ($(b)+(2-0.65,2-0.65)$);
\node at ($(b)+(0,2)-(0.2,0)$) {1};
\node at ($(b)-(0.2,0)$) {2};
\node at ($(b)+(2,0)+(0.2,0)$) {4};
\node at ($(b)+(2,2)+(0.2,0)$) {3};
\draw[line width=3pt] ($(c)+(1,1)$) circle (0.5);
\draw[line width=1pt] (c) -- ($(c)+(0.65,0.65)$);
\draw[line width=1pt] ($(c)+(2,0)$) -- ($(c)+(2-0.65,0.65)$);
\draw[line width=1pt] ($(c)+(0,2)$) -- ($(c)+(0.65,2-0.65)$);
\draw[line width=1pt] ($(c)+(2,2)$) -- ($(c)+(2-0.65,2-0.65)$);
\node at ($(c)+(0,2)-(0.2,0)$) {1};
\node at ($(c)-(0.2,0)$) {3};
\node at ($(c)+(2,0)+(0.2,0)$) {2};
\node at ($(c)+(2,2)+(0.2,0)$) {4};
\draw[line width=1pt] (d) -- ($(d)+(0,2)$);
\draw[line width=1pt] ($(d)+(2,0)$) -- ($(d)+(2,2)$);
\node at ($(d)+(0,2)-(0.2,0)$) {1};
\node at ($(d)-(0.2,0)$) {2};
\node at ($(d)+(2,0)+(0.2,0)$) {3};
\node at ($(d)+(2,2)+(0.2,0)$) {4};
\draw[line width=1pt] (e) -- ($(e)+(0.9,0.9)$);
\draw[line width=1pt] ($(e)+(1.1,1.1)$) -- ($(e)+(2,2)$);
\draw[line width=1pt] ($(e)+(2,0)$) -- ($(e)+(0,2)$);
\node at ($(e)+(0,2)-(0.2,0)$) {1};
\node at ($(e)-(0.2,0)$) {2};
\node at ($(e)+(2,0)+(0.2,0)$) {3};
\node at ($(e)+(2,2)+(0.2,0)$) {4};
\draw[line width=1pt] (f) -- ($(f)+(2,0)$);
\draw[line width=1pt] ($(f)+(0,2)$) -- ($(f)+(2,2)$);
\node at ($(f)+(0,2)-(0.2,0)$) {1};
\node at ($(f)-(0.2,0)$) {2};
\node at ($(f)+(2,0)+(0.2,0)$) {3};
\node at ($(f)+(2,2)+(0.2,0)$) {4};
\end{tikzpicture}
\caption{Diagrammatic representation of four-point colour structures. On the first line we have the single-trace structures $\text{Tr}(1234)$, $\text{Tr}(1243)$, $\text{Tr}(1324)$, while the second line depicts the double-traces $\Tr(12)\Tr(34)$, $\Tr(13)\Tr(24)$, $\Tr(14)\Tr(23)$. The thin lines denote external adjoint indices, while the bold circles stand for contractions of fundamental indices.
}\label{fig:colour_diagrams}
\end{figure}

On the other hand, all colour structures obtained this way must be consistent with the generic colour decomposition familiar from loop amplitudes in flat space. In particular, as is obvious from a diagrammatic approach, the colour basis for gluonic loop-amplitudes is given in terms of traces over fundamental generators of the gauge group. At tree level, this yields the decomposition into colour-ordered amplitudes shown in equation \eqref{eq:M1_colour_ordered}, featuring the single-trace colour structures $\Tr(1234)$, $\Tr(1243)$, and $\Tr(1324)$.\footnote{The fact that there are only three independent single-trace structures is due to the antisymmetry of $SO(N)$ generators. Other gauge groups generically have six independent single-traces, but up to this modification the following statements regarding the loop-level colour decomposition remain valid.}
At loop level however, there are additional contributions from higher-trace terms, see e.g. \cite{BERN1991389}. In the case of four-point amplitudes, the only possible extra structures are double-traces, which we denote by
\begin{align}
	\Tr(12)\Tr(34)\equiv\Tr(T^{I_1}T^{I_2})\Tr(T^{I_3}T^{I_4})\,,
\end{align}
et cetera for the other permutations. In total, there are three such double-trace terms, and their decomposition into $SO(8)$ irreps reads
\begin{align}\label{eq:dtr_decomp}
\begin{split}
	\Tr(12)\Tr(34)&=\delta^{I_1I_2}\delta^{I_3I_4}=\big\{28,0,0,0,0,0,0\big\},\\
	\Tr(13)\Tr(24)&=\delta^{I_1I_3}\delta^{I_2I_4}=\big\{1,1,1,1,1,-1,-1\big\},\\
	\Tr(14)\Tr(23)&=\delta^{I_1I_4}\delta^{I_2I_3}=\big\{1,1,1,1,1,1,1\big\}.
\end{split}
\end{align}
A diagrammatic depiction of these colour structures is given in Figure \ref{fig:colour_diagrams}.

With this in place, the colour structure of loop-level gluon amplitudes takes the form
\begin{align}\label{eq:M_loop_colour}
\begin{split}
	\mathcal{M}^{(k,m)} &= \quad\mathcal{M}^{(k,m)}(1234)\Tr(1234) + (2\text{ permutations})\\
	&\quad+\mathcal{M}^{(k,m)}(12;34)\Tr(12)\Tr(34) + (2\text{ permutations})\,,
\end{split}
\end{align}
and is valid for any loop order $k-1$ and at any order in the $1/\lambda$ expansion. As such, the full amplitude is entirely determined in terms of the single- and double-trace partial amplitudes, $\mathcal{M}^{(k,m)}(1234)$ and $\mathcal{M}^{(k,m)}(12;34)$, and all other partial amplitudes are easily recovered by crossing.

The applicability of the above colour decomposition to AdS amplitudes can already be seen from the leading logs computed in Section \ref{sec:leading_log_one-loop}. Their colour structure arises from the multiplication of tree-level colour vectors, given by the single-trace terms \eqref{eq:tr_decomp}. One notes that these vectors are such that
\begin{enumerate}[label=(\alph*)]
\item the irreps $\mathbf{35'}$ and $\mathbf{35''}$ appear symmetrically, and
\item they have vanishing contribution to the irrep $\mathbf{350}$\,.
\end{enumerate}
Now, since both properties are preserved under addition and multiplication, the loop-level leading log's will have the same feature. In particular, the above two conditions carve out a 5-dimensional subspace within the full 7-dimensional space of $SO(8)$ irreps. One can check that the exact same subspace is spanned by the colour structures
\begin{align}\label{eq:basis_leading_log}
	\big\{\text{Tr}(1234),\,\text{Tr}(1243),\,\text{Tr}(1324),\,\Tr(12)\Tr(34),\,\Tr(13)\Tr(24)+\,\Tr(14)\Tr(23)\big\},
\end{align}
which therefore constitute a complete basis (in colour space) for the leading log. The crossing-completion of the above elements then leads to the general decomposition \eqref{eq:M_loop_colour}.

Finally, as a small digression, let us consider the leading log of the field theory amplitudes $\mathcal{H}_\mathbf{a}^{(k,0)}$. As shown in \cite{Huang:2023oxf}, the relevant colour structures are simply given by $s$-channel planar ladder diagrams at loop order $k-1$, defined as (see Figure 2 in \cite{Huang:2023oxf})
\begin{align}
	\mathbf{p}_{s_1}^{(k-1)} = (-c_t)^k\,,\qquad \mathbf{p}_{s_2}^{(k-1)} = (-c_t)^{k-1}\,c_u\,,
\end{align}
with the two being related by $1\leftrightarrow2$ exchange. According to the above arguments, these colour structures can be written in terms of the basis \eqref{eq:basis_leading_log}. Taking combinations of definite parity, we indeed find the all-orders relations
\begin{align}
\begin{split}
	21\big(\mathbf{p}_{s_1}^{(k-1)}+\mathbf{p}_{s_2}^{(k-1)}\big) &= 14\big(2^{k+1}+(-1)^k\big)\big[\Tr(1234)+\Tr(1243)-2\,\Tr(1324)\big]\\
	&~~\,+\big(9\cdot6^k-7\cdot2^k-2(-1)^k\big)\,\Tr(12)\Tr(34)\\
	&~~\,+14\big(2^k-(-1)^k\big)\big[\Tr(13)\Tr(24)+\Tr(14)\Tr(23)\big],
\end{split}
\end{align}
and
\begin{align}
	\mathbf{p}_{s_1}^{(k-1)}-\mathbf{p}_{s_2}^{(k-1)} = 2\cdot3^k\big[\Tr(1234)-\Tr(1243)\big].
\end{align}

\section{One-loop amplitudes to order $\lambda^{-2}$}\label{sec:oneloopamp}
With the results for the leading logs at hand, we now turn to the reconstruction of the full amplitudes. As we will see, the string corrected one-loop amplitudes take a very simple when written in their Mellin space representation. Our strategy is thus to start directly from an ansatz in Mellin space and constrain the Mellin amplitudes by imposing $(i)$ matching the leading log computed in the previous section, and $(ii)$ crossing symmetry. This simple procedure fixes the entire Mellin amplitude $\mathcal{M}^{(2,m)}(s,t)$ up to a finite number of polynomial ambiguities. After presenting explicit results up to order $\lambda^{-2}$, we also comment on the position space representation, see Section \ref{sec:position_space}.

\subsection{Mellin space algorithm}\label{sec:Mellin_algorithm}
To motivate the ansatz for the Mellin amplitude, it is useful to recall the structure of the one-loop leading logs described in Section \ref{sec:leading_log_one-loop}. Due to the truncation in spin of the string-corrected anomalous dimensions, the leading log contains at most a single power of $\log(V)$. In the Mellin amplitude, we are thus instructed to include only \textit{single} poles in $s$, which together with the Gamma functions in the Mellin transform \eqref{eq:Mellin_trafo_2222} produce the desired $\log^2(U)$ power.\footnote{Similar one-loop computations with spin truncated spectra have been performed for $\phi^4$ theory in AdS$_5$ \cite{Aharony:2016dwx}, and string corrections to super-graviton scattering in AdS$_5\times$S$^5$, see references \cite{Alday:2018kkw,Drummond:2020uni,Aprile:2022tzr}. This is in contrast to the field theory amplitude $\mathcal{M}^{(2,0)}(s,t)$, where \textit{simultaneous} simple poles in $s$ and $t$ are required to match the leading log, see \cite{Alday:2021ajh}.}
A crossing-complete ansatz for the Mellin amplitude then reads
\begin{align}\label{eq:ansatz_M}
	\mathcal{M}^{(2,m)}_{\mathbf{a}}(s,t)=\underbrace{f_\mathbf{a}^{(s)}(s,t)\,\widetilde{\psi}(-s)}_{\mathcal{M}^{(2,m)}_{\mathbf{a}}\vert_s}+\underbrace{f^{(t)}_\mathbf{a}(s,t)\,\widetilde{\psi}(-t)}_{\mathcal{M}^{(2,m)}_{\mathbf{a}}\vert_t}+\underbrace{f_{\mathbf{a}}^{(u)}(s,t)\,\widetilde{\psi}(-u)}_{\mathcal{M}^{(2,m)}_{\mathbf{a}}\vert_u}+\underbrace{p^{(m)}_\mathbf{a}(s,t)}_{\text{ambiguities}}\,,
\end{align}
where by $\mathcal{M}^{(2,m)}_{\mathbf{a}}\vert_{s,t,u}$ we denote the $s$-, $t$- and $u$-channel sub-amplitudes. These are each given by a shifted digamma function $\widetilde{\psi}$ defined by $\widetilde{\psi}(x)\equiv\psi(x)+\gamma_{E}$,\footnote{
	As observed in \cite{Drummond:2019hel}, this shift removes the appearance of the unphysical Euler-Mascheroni constant $\gamma_{E}$ in the position space representation, i.e. when inverting the Mellin transform to pass to the position space correlator $\mathcal{H}$. For more details on position space, see Section \ref{sec:position_space}.}
which provides the desired set of simple poles, multiplied by a corresponding coefficient function $f^{(s,t,u)}_\mathbf{a}(s,t)$. As expected from the spin-truncated nature of the leading log, these coefficient functions turn out to be polynomials in the Mellin variables of degree $m$. The last term represents contributions with no poles, which our procedure is not able to fix. We will comment on the nature of these ambiguities further below.

Our algorithm to fix the undetermined polynomials $f^{(s,t,u)}_\mathbf{a}(s,t)$ in the above ansatz is then given by the following two-step process:
\begin{enumerate}[label=\textbf{(\arabic*)}]
\item \textbf{Matching the leading log:} Plug the ansatz \eqref{eq:ansatz_M} into the Mellin integral and focus on the triple poles in $s$ and double poles in $t$. Performing the residue calculation, one obtains the $\log^2(U)\log(V)$ contribution in a power series around small $U$ and $V$, which one matches against the leading-log prediction. This fixes the $s$-channel polynomials $f_\mathbf{a}^{(s)}(s,t)$.

At this stage, it is furthermore possible to perform the following \textit{consistency check}: with the $s$-channel sub-amplitude fixed, one can verify that in fact the entire $\log^2(U)$ part, i.e. also the coefficient of $\log^0(V)$, is correctly reproduced. This is a non-trivial check on the validity of the ansatz \eqref{eq:ansatz_M}.
\item \textbf{Imposing crossing symmetry:} The remaining polynomials $f^{(t)}_\mathbf{a}(s,t)$ and $f^{(u)}_\mathbf{a}(s,t)$, which do not contribute to the leading log, are then fixed by imposing full crossing symmetry of the amplitude. By means of the crossing equations \eqref{eq:crossing_M}, we simply have
\begin{align}\label{eq:crossing_f}
	f^{(t)}_\mathbf{a}(s,t) = (F_t)_\mathbf{a}^{~\mathbf{b}}\,f^{(s)}_\mathbf{b}(t,s)\,,\quad f^{(u)}_\mathbf{a}(s,t) = (F_u)_\mathbf{a}^{~\mathbf{b}}\,f^{(s)}_\mathbf{b}(u,t)\,.
\end{align}
\end{enumerate}
In this way, knowledge of the leading log entirely fixes all polar parts of the Mellin amplitude.

\paragraph{Ambiguities:} As already indicated in \eqref{eq:ansatz_M}, our results for the one-loop Mellin amplitudes will suffer from certain ambiguities, denoted by $p^{(m)}_\mathbf{a}(s,t)$. This is because it is always possible to add terms with no poles which can not be fixed by the above procedure. Such regular terms are given by polynomials in the Mellin variables, and can therefore be interpreted as contact-term ambiguities to the one-loop amplitudes. Within our algorithm, they are only constrained by step \textbf{(2)}, that is crossing symmetry. However, an additional constraint comes from considering the flat-space limit (further discussed in Section \ref{sec:flat-space-limit}), which sets an upper bound on the polynomial degree correlated with the order in $1/\lambda$:
\begin{align}\label{eq:amb_degree}
	O(\lambda^{-\frac{m}{2}}):\quad\text{deg}\big(p^{(m)}_\mathbf{a}(s,t)\big) \leq m\,.
\end{align}

Note that these polynomial ambiguities are exactly of the form of tree-level string corrections, c.f. Section \ref{sec:tree_level_review}, albeit with (possibly) different coefficients. In fact, the one-loop ambiguity $p^{(m)}_\mathbf{a}(s,t)$ at order $\lambda^{-\frac{m}{2}}$ can be thought of as the genus-one completion of the tree-level amplitude $\mathcal{M}^{(1,m+2)}(s,t)$, where the shift in the power of $\lambda$ is due to the relation $g_s\sim\lambda/N$, recall equation \eqref{eq:dictionary}.

\subsection{Results in Mellin space}
While the algorithm outlined above is in principle valid to any order in $1/\lambda$, it crucially relies on the knowledge of the leading log. As described in Section \ref{sec:leading_log_one-loop}, the currently available tree-level data allows us to compute the leading log unambiguously at orders $\lambda^{-1}$ and $\lambda^{-\frac{3}{2}}$, while at order $\lambda^{-2}$ one already has four free parameters. To limit the proliferation of free parameters, we demonstrate our algorithm in the following by constructing the first three one-loop string corrections. Note that, when presenting the explicit results for $\mathcal{M}^{(2,2)}(s,t)$, $\mathcal{M}^{(2,3)}(s,t)$ and $\mathcal{M}^{(2,4)}(s,t)$, we will omit writing out the ambiguities. It is understood that the results are only fixed up to the polynomials $p^{(m)}_\mathbf{a}(s,t)$, which are constrained by crossing symmetry and the bound \eqref{eq:amb_degree}.

\subsubsection{Order $\lambda^{-1}$}\label{sec:result_M22}
Following step \textbf{(1)} of our algorithm, we start by matching the Mellin space ansatz \eqref{eq:ansatz_M} against the leading log at this order, which we recall is given in \eqref{eq:llog_2}. This determines the $s$-channel sub-amplitude to be
\begin{align}\label{M221}
	\mathcal{M}^{(2,2)}_\mathbf{a}\vert_s = \begin{pmatrix} 90 \f (s)\\ 14\f(s)\\ -4\f(s)\\ -4\f(s)\\ -\f(s)\\\hline 0 \\ 0 \end{pmatrix},\quad \f(s)=-32\zeta_2(5s^2+7s+3)\,\widetilde{\psi}(-s)\,.
\end{align}
Note that the polynomial $\f(s)$ does not depend on $t$, in accordance with having only spin-0 contributions to the leading log.
Proceeding to step \textbf{(2)}, the completion of the above $s$-channel sub-amplitude is obtained by adding the crossing images by means of equation \eqref{eq:crossing_f}. This already yields the final result (written in the basis of irreps):
\begin{align}\label{eq:M22_result}
	\mathcal{M}_\mathbf{a}^{(2,2)}(s,t) = \begin{pmatrix}
				90\f(s)\\ 14\big[\f(s)+\f(t)+\f(u)\big]\\ -4\big[\f(s)+\f(t)+\f(u)\big]\\ -4\big[\f(s)+\f(t)+\f(u)\big]\\ -\frac{1}{2}\big[2\f(s)-7\f(t)-7\f(u)\big]\\\hline \frac{15}{2}\big[\f(t)-\f(u)\big]\\ 3\big[\f(t)-\f(u)\big]
				\end{pmatrix}.
\end{align}
As argued in Section \ref{sec:colour_structure_loops}, we can express the above result in the more familiar one-loop colour basis \eqref{eq:M_loop_colour}. Performing this change of basis using equations \eqref{eq:tr_decomp} and \eqref{eq:dtr_decomp}, the single- and double-trace partial amplitudes equivalent to \eqref{eq:M22_result} simply read
\begin{align}
\begin{split}
	\mathcal{M}^{(2,2)}(1234) &= 5\f(s)+5\f(t)+2\f(u)\,,\\
	\mathcal{M}^{(2,2)}(12;34) &= 2\f(s)-\f(t)-\f(u)\,,\\
\end{split}
\end{align}
and the remaining partial amplitudes are easily obtained by crossing.

\subsubsection{Order $\lambda^{-3/2}$}\label{sec:result_M23}
Proceeding as before, we first match the leading which at this order is given in \eqref{eq:llog_3}. Comparing against the ansatz in Mellin space, we find
\begin{align}\label{M231}
	\mathcal{M}^{(2,3)}_\mathbf{a}\vert_s = \begin{pmatrix} 36\g(s)\\ 4\g(s)\\ 4\g(s)\\ 4\g(s)\\ \g(s)\\\hline 3 g(s,t)\\ 0 \end{pmatrix},
\end{align}
where
\begin{align}
\begin{split}
	\g(s)&=-256\zeta_3(15s^3+25s^2+20s+6)\,\widetilde{\psi}(-s)\,,\\
	g(s,t)&=-\frac{768\zeta_3}{5}(15s^2+25s+12)(s+2t+3)\,\widetilde{\psi}(-s)\,.
\end{split}
\end{align}
In the symmetric channels, we again have a polynomial with no $t$-dependence, reflecting the spin-0 truncation of the leading log. On the other hand, the polynomial in the antisymmetric irrep $\mathbf{28}$ is of degree one in $t$, consistent with the exchange of a spin-1 operator. Moreover, in accordance with invariance under $1\leftrightarrow2$ exchange of the $s$-channel sub-amplitude, the polynomial $g(s,t)$ has the symmetry 
\begin{align}
	g(s,t)=-g(s,u)\,.
\end{align}
To obtain the full amplitude, we use again \eqref{eq:crossing_f}. This yields the result
\begin{align}\label{eq:M23_final}
	\mathcal{M}^{(2,3)}(s,t) = \begin{pmatrix}
				3\big[12 \g(s) + 9 \g(t) + 9 \g(u) + g(t,s) - g(u,t)\big]\\ 4 \g(s) + \g(t) + \g(u) + g(t, s) - g(u,t)\\ 4 \g(s) + \g(t) + \g(u) + g(t, s) - g(u,t)\\ 4 \g(s) + \g(t) + \g(u) + g(t, s) - g(u,t)\\ \frac{1}{2}\big[2 \g(s) + 5 \g(t) + 5 \g(u) - g(t,s) + g(u,t)\big]\\\hline \frac{3}{2}\big[3 \g(t) - 3 \g(u) + 2 g(s,t) + g(t,s) + g(u,t)\big]\\ 0
				\end{pmatrix},
\end{align}
where the last entry (corresponding to the irrep $\mathbf{350}$) being zero is due to a cancellation between the $t$- and $u$-channel sub-amplitudes.

Recasting the above into the standard one-loop basis \eqref{eq:M_loop_colour}, the two independent partial amplitudes read
\begin{align}
\begin{split}
	\mathcal{M}^{(2,3)}(1234) &= \g(s)+\g(t)-2\g(u)+g(s,t)+g(t,s)\,,\\
	\mathcal{M}^{(2,3)}(12;34) &= \g(s)+\g(t)+\g(u)\,.
\end{split}
\end{align}
Interestingly, we note that the double-trace partial amplitude turns out to be fully crossing symmetric, i.e. one has $\mathcal{M}^{(2,3)}(12;34)=\mathcal{M}^{(2,3)}(13;24)=\mathcal{M}^{(2,3)}(14;23)$. This observation is in fact related to a special property of the colour structure at this order: 

As mentioned above equation \eqref{eq:M13_c}, both the tree-level field theory amplitude as well as the $\lambda^{-3/2}$ correction can be written in terms of the tree-level colour structures $c_{s,t,u}$. The colour structure of the one-loop leading log, obtained by gluing these tree-level structures, is therefore given in terms of one-loop box diagrams in colour space \cite{Alday:2021ajh}, defined by\footnote{The decomposition of $d_{st,su,tu}$ into the one-loop basis \eqref{eq:M_loop_colour} reads
\begin{align}\label{eq:one-loop_identities}
\begin{split}
	d_{st} &= \Tr(12)\Tr(34)+\Tr(13)\Tr(24)+\Tr(14)\Tr(23)+4\Tr(1234)-2\Tr(1243)-2\Tr(1324)\,,\\
	d_{su} &= \Tr(12)\Tr(34)+\Tr(13)\Tr(24)+\Tr(14)\Tr(23)-2\Tr(1234)+4\Tr(1243)-2\Tr(1324)\,,\\
	d_{tu} &= \Tr(12)\Tr(34)+\Tr(13)\Tr(24)+\Tr(14)\Tr(23)-2\Tr(1234)-2\Tr(1243)+4\Tr(1324)\,.
\end{split}
\end{align}}
\begin{align}\label{eq:d_structures}
\begin{split}
	d_{st} &= f^{JI_1K}f^{KI_2L}f^{LI_3M}f^{MI_4J} =\big\{36, 4, 4, 4, 1, 9, 0\big\},\\
	d_{su} &= f^{JI_1K}f^{KI_2L}f^{LI_4M}f^{MI_3J} =\big\{36, 4, 4, 4, 1, -9, 0\big\},\\
	d_{tu} &= f^{JI_1K}f^{KI_3L}f^{LI_2M}f^{MI_4J} =\big\{18, -2, -2, -2, 4, 0, 0\big\}.
\end{split}
\end{align}
Thus, instead of using the one-loop basis \eqref{eq:M_loop_colour}, we can alternatively write the previous result \eqref{eq:M23_final} for $\mathcal{M}^{(2,3)}$ in the form
\begin{align}\label{eq:M23_d_structures}
	\mathcal{M}^{(2,3)}(s,t) = \mathcal{M}_{st}\,d_{st}+\mathcal{M}_{su}\,d_{su}+\mathcal{M}_{tu}\,d_{tu}\,,
\end{align}
with $\mathcal{M}_{st}$ given by
\begin{align}
\begin{split}
	\mathcal{M}_{st} &= \frac{f_3(s)+f_3(t)}{2} + \frac{g(s,t)+g(t,s)}{6}\\
				     &= \frac{1}{10}\big(90 s^3+30 s^2 t+195 s^2+50 s t+187 s+24 t+66\big)\,\widetilde{\psi}(-s) + (s\leftrightarrow t)\,.
\end{split}
\end{align}
As required by crossing symmetry (and suggested by the notation), this term is manifestly symmetric under $s\leftrightarrow t$. Furthermore, as is the case for the one-loop field theory amplitude $\mathcal{M}^{(2,0)}$ from \cite{Alday:2021ajh}, this rewriting has the property that $\mathcal{M}_{st}$ does not contain any poles in $u$, but only in $s$ and $t$. Finally, let us stress that this rewriting is not possible for any other one-loop string correction discussed here, as it relies on the special properties of the tree-level $\lambda^{-3/2}$ amplitude \eqref{eq:M13_c}.

\subsubsection{Order $\lambda^{-2}$}
As in the derivation of the leading log at this order, it is instructive to split the result for $\mathcal{M}_\mathbf{a}^{(2,4)}(s,t)$ into two parts,
\begin{align}
	\mathcal{M}_\mathbf{a}^{(2,4)}(s,t) \equiv \mathcal{M}_\mathbf{a}'^{(2,4)}(s,t) + \mathcal{M}_\mathbf{a}''^{(2,4)}(s,t)\,,
\end{align}
and we recall that the first term comes from squaring the order $\lambda^{-1}$ tree-level anomalous dimension, while the second term comes from the product of the field-theory and the order $\lambda^{-2}$ anomalous dimensions.

\paragraph{Part 1:}
The first and simpler part leads to the following $s$-channel sub-amplitude,
\begin{align}
	\mathcal{M}'^{(2,4)}_\mathbf{a}\vert_s = \begin{pmatrix} 225\h(s)\\ 49\h(s)\\ 4 \h(s)\\ 4\h(s)\\ \h(s)\\\hline 0 \\ 0 \end{pmatrix},\quad \h(s)=-\frac{1536(\zeta_2)^2}{5}(35s^4+50s^3+55s^2+28s+6)\,\widetilde{\psi}(-s)\,,
\end{align}
whose structure is unsurprisingly similar to the order $\lambda^{-1}$ result, c.f. equation \eqref{M221}.
After crossing-completion, we find that the partial amplitudes are given by
\begin{align}
\begin{split}
	\mathcal{M}'^{(2,4)}(1234) &= 16\h(s)+16 \h(t)-2 \h(u)\,,\\
	\mathcal{M}'^{(2,4)}(12;34) &= 4\h(s)+ \h(t)+ \h(u)\,.\\
\end{split}
\end{align}

\paragraph{Part 2:}
The analogous expressions for the second part turn out to be quite lengthy, which is due to the presence of the four free parameters in the leading log. We will therefore write out explicitly only the quartic terms of the coefficient polynomials, and the full expressions are recorded in the ancillary file. Having said this, the $s$-channel sub-amplitude is of the form
\begin{align}\label{Ma42}
	\mathcal{M}''^{(2,4)}_\mathbf{a}\vert_{s}= -\frac{7\cdot2^9\,\zeta_2^2}{5}\begin{pmatrix} 27 s^2(221 s^2 +7(t^2+u^2 ))+\ldots\\  s^2(893 s^2 +31(t^2+u^2 ))+\ldots\\ -2s^2(71 s^2 +7(t^2+u^2 ))+\ldots\\ -2s^2(71 s^2 +7(t^2+u^2 ))+\ldots\\ -\frac{1}{2}s^2(71 s^2 +7(t^2+u^2 ))+\ldots\\\hline -81 s^3 (t-u)+\ldots \\ 0 \end{pmatrix}\widetilde{\psi}(-s)\,.
\end{align}
Computing the crossing-completion and extracting the partial amplitudes then yields
\begin{align}
\begin{split}
	\mathcal{M}''^{(2,4)}(1234) = -\frac{7\cdot2^9\,\zeta_2^2}{5} & \Big[\big(294 s^4 + 23 s^2 t^2 -31 s^3 t +\ldots\big)\,\widetilde{\psi}(-s)\\
&	+ \big( 294 t^4 -31 s t^3 + 23  s^2 t^2  +\ldots\big)\,\widetilde{\psi}(-t)\\
&  + \big( 78 (s^2+t^2) u^2 + 142 s t u^2 +\ldots\big)\,\widetilde{\psi}(-u)\Big],\\[3pt]
	\mathcal{M}''^{(2,4)}(12;34) = -\frac{7\cdot2^9\,\zeta_2^2}{5} & \Big[ \big( 141 s^2 (t^2 + u^2) + 274 s^2 t u +\ldots\big)\,\widetilde{\psi}(-s)\\
&	-\big(  39  t^4 +7 t^2 u (t+u) +\ldots\big)\,\widetilde{\psi}(-t)\\
&   -\big(  39  u^4 +7 t u^2 (t+u) +\ldots\big)\,\widetilde{\psi}(-u) \Big].
\end{split}
\end{align}
As mentioned before, these expressions contain the four free parameters $a_2$, $b_1$, $e_1$ and $f_1$ (hidden in the subleading terms), which are inherited from the tree-level amplitude $\mathcal{M}^{(1,4)}$ given in \eqref{eq:M14}. In particular, the free parameters $a_2$ and $b_1$ are by construction proportional to the one-loop $\lambda^{-1}$ and $\lambda^{-3/2}$ amplitudes given in Sections \ref{sec:result_M22} and \ref{sec:result_M23}. On the other hand, the other two parameters $e_1$ and $f_1$ are linearly independent from the amplitudes at previous orders. As such, they can be seen as the imprint of the $\langle22pp\rangle$ family of correlators, which was used in the construction of the leading log.

\subsection{The position-space representation}\label{sec:position_space}
Before discussing the flat-space limit, let us also comment on the form of the position space correlators $\mathcal{H}^{(2,n)}(x,\xb)$. As we will see, considering the position space representation allows us to observe a remarkable simplification, namely that it is possible to pull out the differential operator $\dfour$ \textit{twice}, a property which we find more difficult to spot in the Mellin space formulation of the previous section.

In principle, the position space correlators can be computed from the Mellin amplitudes by evaluating the double contour integral \eqref{eq:Mellin_trafo_2222}. However, in practice it is difficult to perform these integrals analytically, and it is better to start from a suitable ansatz directly in position space instead. This strategy has been successfully applied to the case of one-loop string corrections to supergraviton scattering in AdS$_5\times$S$^5$, where the Mellin space amplitudes are also given by a digamma function with polynomial coefficients. As shown in \cite{Drummond:2019hel}, the corresponding position space ansatz consists of single-valued multiple polylogarithms (SVMPL's) of transcendental weight up to 3, built from the alphabet of letters $\{x,\xb,1-x,1-x,x-\xb\}$. The relevant set of functions contains 10 elements, which we denote by $\mathcal{Q}_i(x,\xb)$. Listed in order of increasing transcendental weight, these functions read:
\begin{align}\label{eq:transcendental_basis}
\begin{split}
	\bullet~\text{weight 0:}&\qquad 1\,,\\[3pt]
	\bullet~\text{weight 1:}&\qquad \log(U),\,\log(V)\,,\\[3pt]
	\bullet~\text{weight 2:}&\qquad \phi^{(1)}(x,\xb),\,\log^2(U),\,\log(U)\log(V),\,\log^2(V)\,,\\[3pt]
	\bullet~\text{weight 3:}&\qquad f^{(3)}(x,\xb),\,\log(U)\,\phi^{(1)}(x,\xb),\,\log(V)\,\phi^{(1)}(x,\xb)\,,
\end{split}
\end{align}
where $\phi^{(1)}(x,\xb)$ is the well known one-loop box integral given by
\begin{align}
	\phi^{(1)}(x,\xb) = \big(\text{log}(1-x)-\log(1-\xb)\big)\log(x\xb)+2\big(\Li_2(x)-\Li_2(\xb)\big).
\end{align}

The function $f^{(3)}(x,\xb)$ is the only basis element which contains the letter $x-\xb$.\footnote{The precise form of this function is somewhat lengthy to write out, so we will not give it here. An explicit expression in terms of multiple polylogarithms can be found in the ancillary file, where we used the conventions of the package \texttt{PolyLogTools} \cite{Duhr:2019tlz}. For a full characterisation of $f^{(3)}(x,\xb)$ and its properties see also Section 2.3 and Appendix A of \cite{Drummond:2019hel}.} As explained in \cite{Drummond:2022dxw}, the presence of this letter leads to an extra logarithmic divergence $\log(x-\xb)$ in the bulk point limit. Physically, this signals the presence of a scale-dependent term due to a logarithmic divergence in the flat-space amplitude. This also explains why $f^{(3)}(x,\xb)$ appears in the one-loop field theory correlator of \cite{Huang:2023oxf}: the eight-dimensional one-loop box integral in flat space is indeed logarithmically divergent! On the other hand, this function can \textit{not} appear in the one-loop supergravity correlator on AdS$_5\times$S$^5$ \cite{Aprile:2017bgs,Aprile:2019rep}, since the scale-dependent logarithmic term of the ten-dimensional box integral precisely cancels after adding up the three crossing orientations.\footnote{However, what does not cancel is the quadratic divergence. In the AdS amplitude, this renormalisation term is reflected by the presence of a contact-term ambiguity -- much like the finite spin ambiguities which our bootstrap procedure for the one-loop string corrections is not able to fix.}

The presence of $f^{(3)}(x,\xb)$ is expected also for a different reason. In ref. \cite{Heckelbacher:2022fbx}, the authors evaluate the one-loop bubble diagram in $\phi^4$ theory on AdS$_4$. Upon closer inspection,\footnote{We thank Paul Heslop and Arthur Lipstein for discussion on this point.} we find that their result is simply related to the function $f^{(3)}(x,\xb)$!
Naively, we expect the one-loop string corrections considered in this paper to be somewhat related to a one-loop scalar bubble diagram, \textit{a fortiori} given that tree-level correlators in AdS$_5\times$S$^3$ can be obtained from quartic scalar interactions in an AdS$_5\times$S$^3$ bulk \cite{Glew:2023wik}. 
Even though the space-time dimensions differ, this provides an additional physical motivation why this function appears in the position space representation.

\paragraph{The position-space calculation:}
We are now ready to state the ansatz for the position space correlators $\mathcal{H}^{(2,n)}(x,\xb)$. Within each irrep $\mathbf{a}$, each of the above basis functions $\mathcal{Q}_i(x,\xb)$ is multiplied by a rational coefficient function:
\begin{align}\label{eq:ansatz_H2m}
	\mathcal{H}_\mathbf{a}^{(2,m)}(x,\xb) = \sum_{i=1}^{10}\frac{q_\mathbf{a}^{(i)}(x,\xb)}{U^2\,(x-\xb)^{d_i}}\,\mathcal{Q}_i(x,\xb)\,,
\end{align}
with denominator powers $d_i=d$ for antisymmetric functions $\mathcal{Q}_i(x,\xb)$, while for symmetric functions we have $d_i=d-1$. The maximal denominator power $d$ is then correlated with the order in $1/\lambda$: at order $\lambda^{-\frac{m}{2}}$, one has $d=2m+9$. The polynomials $q_\mathbf{a}^{(i)}(x,\xb)$ are of the form
\begin{align}\label{eq:p_coefficients}
	q_\mathbf{a}^{(i)}(x,\xb)=\sum_{j=0}^{d_i}\sum_{k=j}^{d_i} c_{\mathbf{a},j,k}^{(i)}\big(x^j\xb^k+x^k\xb^j\big),
\end{align}
where the coefficients $c_{\mathbf{a},j,k}^{(i)}$ denote the undetermined parameters of the ansatz.

We then constrain these parameters by imposing certain consistency conditions. In analogy with the Mellin space algorithm of Section \ref{sec:Mellin_algorithm}, we impose matching of the leading log and crossing symmetry. In addition, a further constraint comes from demanding absence of unphysical poles at $x=\xb$, which are introduced by the $(x-\xb)$ denominator powers in the ansatz \eqref{eq:ansatz_H2m}. Note that in the Mellin space formulation, the absence of such poles is already manifest.

Using the above procedure, we have explicitly constructed the correlators up to order $\lambda^{-2}$, that is $\mathcal{H}^{(2,2)}$, $\mathcal{H}^{(2,3)}$ and $\mathcal{H}^{(2,4)}$. By expanding around small $x$ and $1-\xb$, we have verified to high order that they precisely agree with the previously presented results in Mellin space (with all ambiguities set to zero). We now turn to the interesting observation, that the position space correlators can be further simplified by making use of the differential operator $\dfour$.

\paragraph{A remarkable simplification:} By inspecting the results of the above computation, we find that the one-loop string corrections $\mathcal{H}_{\mathbf{a}}^{(2,m)}(x,\xb)$ can be written as $(\dfour)^2$ acting on a simpler object, i.e. one has\footnote{In \cite{Drummond:2022dxw}, a similar feature has been observed for the one-loop $(\ap)^3$ correction to graviton scattering in AdS$_5\times$S$^5$, with $\dfour$ replaced by an eight-order differential operator.}
\begin{align}\label{eq:H2m_simplification}
	\mathcal{H}_{\mathbf{a}}^{(2,m)}(x,\xb) = \big(\dfour\big)^2\,\mathcal{P}_\mathbf{a}^{(2,m)}(x,\xb)\,,
\end{align}
where the pre-correlators $\mathcal{P}_\mathbf{a}^{(2,m)}(x,\xb)$ are of the same form as \eqref{eq:ansatz_H2m} but have a reduced maximal denominator power $d$: instead of $d=2m+9$ for the full correlators $\mathcal{H}_{\mathbf{a}}^{(2,m)}(x,\xb)$, the pre-correlators have $d=2m+1$. As a consequence, also the polynomials \eqref{eq:p_coefficients} in the pre-correlator are of a lower degree, such that the final expressions simplify considerably.

We note that the property \eqref{eq:H2m_simplification} is somewhat surprising, as the one-loop field theory correlator $\mathcal{H}^{(2,0)}(x,\xb)$ allows only for \textit{one} power of $\dfour$ to be extracted \cite{Huang:2023oxf}. Moreover, compared to the one-loop field theory case there are no additional tree-like terms on the RHS of \eqref{eq:H2m_simplification}, and the full correlator is entirely determined by the simpler pre-correlator. Apart from the mentioned simplifications, the pre-correlator further profits from a special property of $(\dfour)^2$: while a generic power of $\dfour$ has only one crossing symmetry, its square is in fact fully crossing symmetric.\footnote{This property has been noticed in \cite{Huang:2023oxf}, where it was an important ingredient to the computation of the two-loop field theory amplitude. Again, this is in complete analogy to the AdS$_5\times$S$^5$ case.}
In our conventions, the symmetry properties of $\dfour$ read
\begin{align}
	\big(\dfour\big)^k\big\vert_{x\to x',\,\xb\to\xb'} = V^3\big(\dfour\big)^k\frac{1}{V^3}\,,\quad \big(\dfour\big)^2\big\vert_{x\to1-x,\,\xb\to1-\xb} = \big(\dfour\big)^2\frac{V^2}{U^2}\,,
\end{align}
where the first symmetry is consistent with the $1\leftrightarrow2$ exchange symmetry of the OPE decomposition and holds for any power $k$. On the other hand, the enhancement to full crossing symmetry is a particular property of the square of $\dfour$. Consequently, the pre-correlator can be made fully crossing symmetric as well, and its transformation properties read
\begin{align}\label{eq:crossing_P}
\begin{split}
	\mathcal{P}^{(2,m)}_\mathbf{a}(x,\xb) &= \frac{1}{V^3}(F_s)_\mathbf{a}^{~\mathbf{b}}\, \mathcal{P}^{(2,m)}_\mathbf{b}(x',\xb') \\
	&= \frac{V^2}{U^2}(F_t)_\mathbf{a}^{~\mathbf{b}}\, \mathcal{P}^{(2,m)}_\mathbf{b}(1-x,1-\xb)\\
	&= \frac{1}{U^5}(F_u)_\mathbf{a}^{~\mathbf{b}}\,\mathcal{P}^{(2,m)}_\mathbf{b}(1/x,1/\xb)\,,
\end{split}
\end{align}
where we used the definition $x'\equiv x/(x-1)$, and similarly for $\xb'$.

Our results for the correlators in position space are recorded in an ancillary file, where we give the explicit results up to order $\lambda^{-2}$ in terms of their pre-correlators (given in the basis \eqref{eq:M_loop_colour} of single- and double-trace colour structures). Note that the differential operator $\dfour$ has a non-trivial kernel, which renders any expression for the pre-correlator inherently ambiguous. In the recorded expressions for $\mathcal{P}^{(2,2)}$, $\mathcal{P}^{(2,3)}$ and $\mathcal{P}^{(2,4)}\equiv\mathcal{P}'^{(2,4)}+\mathcal{P}''^{(2,4)}$ we have arbitrarily fixed such ambiguities.

\paragraph{Ambiguities:} A final comment is in order regarding the contact-term ambiguities, which in the Mellin space representation are described by the polynomials $p_\mathbf{a}^{(m)}(s,t)$, recall \eqref{eq:ansatz_M}. The ambiguities arising in the position space calculation are built from a subset of the transcendental functions \eqref{eq:transcendental_basis}, namely the four basis elements $\big\{\phi^{(1)}(x,\xb),\,\log(U),\,\log(V),\,1\big\}$.\footnote{In fact, they correspond to certain (linear combinations) of $\dbar{}$-functions, with their sum of indices $\leq 2m+12$ at order $\lambda^{-\frac{m}{2}}$.}
We find that these ambiguities are indeed in one-to-one correspondence to those in the Mellin amplitude, and the condition \eqref{eq:amb_degree} on the degree of the Mellin polynomials corresponds to not exceeding the maximal denominator power $d=2m+9$, which is set by the leading log at that order.

\section{The flat-space limit at one loop}\label{sec:flat-space-limit}
In this section, we will analyse the flat-space limit of the previously derived AdS Mellin amplitudes $\mathcal{M}^{(2,m)}$, and compare it to the $s$-channel discontinuity of the genus-one open string amplitude in 8-dimensional flat space.

Recall that at tree-level, the Mellin amplitude and the flat-space amplitude are related via an integral transformation which, in the present case, reads \cite{Penedones:2010ue}:
\begin{equation}\label{flatspacepenedtree}
\mathcal{V}^{(1)}_\mathbf{a}(s,t) = \lim_{R \to \infty}  \pi^4 R^8 \frac{1}{2\pi i} \int d\beta \,  \frac{e^{\beta} }{\beta^4} \mathcal{M}_\mathbf{a}^{(1)} \left(\frac{R^2  s}{4\beta}, \frac{R^2  t}{4\beta} \right),
\end{equation} 
where $\mathcal{V}^{(1)}_\mathbf{a}(s,t)$ is the 8d flat-space Veneziano amplitude, which we recall later in equation \eqref{Venezianoflat}. The notation we employ for the genus expansion follows the same conventions as the AdS amplitude:
\begin{equation}\label{flatfourstring}
\mathcal{V}= s^2 \times \bigl[ g_s \mathcal{V}^{(1)}+ g_s^2 \mathcal{V}^{(2)} +\cdots \bigr],
\end{equation}
where each term admits an $\ap$ expansion:
\begin{align}
\mathcal{V}^{(1)} &= \sum_{m\geq 0} \ap^{(m+2)} \mathcal{V}^{(1,m)},\notag \\
\mathcal{V}^{(2)} &= \sum_{m\geq 0} \ap^{(m+4)} \mathcal{V}^{(2,m)}, 
\end{align}
and so on.
The factor of $s^2$ comes from the polarisation factor $\delta^{8}(\tilde{Q})$ upon restricting to the scalar component of the supermultiplet. The latter can be seen as the flat-space analogue of the factor $\mathcal{I}$  in AdS, c.f. equation \eqref{eq:solution_SCWI}.

Let us now consider genus-one corrections. As observed in \cite{Aprile:2020luw}, the idea is that for large arguments $\psi(-s)$ approaches $\log(-s)$. Thus, we expect to recover the flat-space amplitude, upon integrating the rational function sitting in front of $\psi(-s)$ as in the tree-level flat-space limit prescription. In other words, we expect the following relation between flat and AdS amplitudes to hold for the one-loop discontinuity\footnote{See also \cite{Aprile:2022tzr} where this argument has been applied to genus-one corrections in AdS$_5 \times$S$^5$.}
\begin{equation}\label{flatspacepened1}
\mathcal{V}^{(2)}_\mathbf{a}(s,t)\vert_s = \biggl( \lim_{R \to \infty}  \pi^4 R^8 \frac{1}{2\pi i} \int d\beta \,  \frac{e^{\beta} }{\beta^4} f_\mathbf{a}^{(s)} \left(\frac{R^2  s}{4\beta}, \frac{R^2  t}{4\beta} \right) \biggr)\times \log(-s)\,,
\end{equation}
where $f_\mathbf{a}(s,t)$ is the polynomial sitting in front of $\widetilde{\psi}$,  see equation \eqref{eq:ansatz_M}, and the symbol $\vert_s$ stands for the s-channel part.
It will be convenient to work directly in the basis of irreps.

\paragraph{Order $g_s^2 \ap^6 $:} Applying the prescription \eqref{flatspacepened1} to the order $\lambda^{-1}$ Mellin amplitude from equation \eqref{M221}, we have
\begin{align}\label{flatspacelim2}
\mathcal{M}^{(2,2)}_\mathbf{a}\vert_s \,\rightarrow\, - \frac{16}{3} \zeta_2  \pi^6 \begin{pmatrix} 90\\ 14\\ -4\\ -4\\ -1\\\hline 0 \\ 0 \end{pmatrix}
\times s^2 \log(-s)\,.
\end{align}
where we have used the holographic dictionary \eqref{eq:dictionary} to recast the amplitude in terms of $\ap$ and $g_s$.

\paragraph{Order $g_s^2 \ap^7$:} Similarly, from \eqref{M231} we get
\begin{align}\label{flatspacelim3}
\mathcal{M}^{(2,3)}_\mathbf{a}\vert_s \,\rightarrow\, - \frac{16}{3} \zeta_3  \pi^6 \begin{pmatrix} 36 s\\ 4s\\ 4s\\ 4s\\ s\\\hline \frac{9}{5}(t-u)\\ 0 \end{pmatrix}
\times s^2 \log(-s)\,.
\end{align}

\paragraph{Order $ g_s^2 \ap^8$:} Finally, the two contributions at order $\lambda^{-2}$ yield
\begin{align}\label{flatspacelim41}
\mathcal{M}'^{(2,4)}_\mathbf{a}\vert_s \,\rightarrow\, - \frac{8}{15}  \zeta_2^2  \pi^6 \begin{pmatrix} 225\\ 49\\ 4\\ 4\\ 4\\\hline 1 \\ 0 \end{pmatrix}
\times s^4 \log(-s)\,,
\end{align}
and
\begin{align}\label{flatspacelim42}
\mathcal{M}''^{(2,4)}_\mathbf{a}\vert_s \,\rightarrow\, - \frac{8}{225} \zeta_2^2  \pi^6 \begin{pmatrix} 27(221 s^2 +7(t^2+u^2 ))\\  893 s^2 +31(t^2+u^2 )\\ -2(71 s^2 +7(t^2+u^2 ))\\ -2(71 s^2 +7(t^2+u^2 ))\\ -\frac{1}{2}(71 s^2 +7(t^2+u^2 ))\\\hline -81 s (t-u) \\ 0 \end{pmatrix}
\times s^2 \log(-s)\,.
\end{align}

We are now going to verify through an explicit computation that the above flat-space limits agree with the s-channel discontinuity of the genus-one 8-dimensional amplitude in flat space.

\subsection{The 8-dimensional flat-space open string amplitude}
First, recall that the Veneziano amplitude, (i.e. order $g_s$), reads:
\begin{equation}
\mathcal{V}^{(1)}=  \Tr [1234]  \mathcal{V}^{(1)} (1234)+ \Tr [1243]  \mathcal{V}^{(1)} (1243)+ \Tr [1324]  \mathcal{V}^{(1)} (1324)\,,
\end{equation}
with
\begin{align}\label{Venezianoflat}
\begin{split}
\mathcal{V}^{(1)} (1234) & = -  256 \pi^5 \ap^2  \frac{1}{s t}\,\frac{\Gamma(1-\ap s)\Gamma(1-\ap t)}{\Gamma(1+\ap u)} =  \sum_{m \geq 0} \ap^{(m+2)} \mathcal{V}^{(1,m)} (1234) \\ 
& =  256 \pi^5 \left[- \frac{1}{st}\, \ap^2+\zeta_2\,\ap^4-\zeta_3 u\,\ap^5+\frac{\zeta_2^2}{20} (7s^2+7t^2+u^2)\,\ap^6+\ldots\, \right],
\end{split}
\end{align}  
and analogously for the other colour-ordered amplitudes, where the overall constant is fixed by requiring that the Dirac-Born-Infield action for a D7 brane matches a canonically normalised 8-dimensional Yang-Mills amplitude in the field-theory limit \cite{Behan:2023fqq}. 
Here the flat-space Mandelstam variables obey $s+t+u=0$.

Our goal is to reconstruct the discontinuity of the genus-one amplitude,
order by order in $\ap$, by gluing tree-level partial-wave coefficients.
The computation follows closely the construction of the AdS amplitude, with the conformal block expansion replaced by the usual partial-wave expansion in Gegenbauer polynomials. 
Schematically, the recipe is thus as follows:
\begin{enumerate}[label=\textbf{(\arabic*)}]
\item Decompose the flat-space Veneziano amplitude into projectors order by order in $\ap$.
\item Compute the partial wave coefficients in each channel.
\item Multiply the tree-level partial wave coefficients and resum the expression to obtain the s-channel cut in each channel at the desired order.
\end{enumerate}

Implementing this, we have
\begin{equation}
\mathcal{V}^{I_1 I_2 I_3 I_4} =\sum_{\bf{a}} \mathcal{V}_{\bf{a}}  \mathbb{P}_{\bf{a}}^{I_1 I_2 I_3 I_4}.
\end{equation} 
We then consider the partial wave expansion in each channel:
\begin{equation}
\mathcal{V}_{\bf{a}} = \frac{2 i}{s^{\frac{d-4}{2}}}\sum_{\ell} \left(1-e^{2i\epsilon_{\bf{a},\ell}(s)} \right)P_{\ell}^{\frac{d-3}{2}}(z)\,, \qquad z=1+\frac{2 t}{s}\,,
\end{equation}
with $P_{\ell}^{\frac{d-3}{2}}(z)$ proportional to Gegenbauer polynomials
\begin{equation}
P_{\ell}^{\frac{d-3}{2}}(z)=2^{2d-5}\pi^{\frac{d-3}{2}}(d+2\ell-3)\Gamma \left[\frac{d-3}{2} \right]C_{\ell}^{\frac{d-3}{2}}(z)\,,
\end{equation}
and the sum running over even (odd) spins for symmetric (antisymmetric) channels. In the following we will restrict to $d=8$.
The decomposition is clearly also valid order by order in $\alpha'$, with the phase shift admitting an analogous double expansion as the amplitude:
\begin{equation}
\epsilon_{\bf{a},\ell}=g_s \epsilon_{\bf{a},\ell}^{(1)}+g_s^2 \epsilon_{\bf{a},\ell}^{(2)}+\ldots, \qquad
 \epsilon_{\bf{a},\ell}^{(1)}=\sum_{m\geq 0} \ap^{(m+2)} \epsilon_{\bf{a},\ell}^{(1,m)},
\end{equation}
and so on.

In the field-theory limit we have,
\begin{equation}
\mathcal{V}_{\bf{a}}^{(1,0)}=
256 \pi^5  \frac{1}{s t u} \begin{pmatrix}
 3 s \\
  s \\
  s \\
  s \\
 -\frac{1}{2 }s  \\
 \hline
 \frac{3}{2}(t-u)  \\
 0  \\
\end{pmatrix},
\end{equation}
and, accordingly,\footnote{The formula agrees with the result of \cite{Caron-Huot:2018kta,Guerrieri:2021ivu} upon restricting to $d=8$.}\textsuperscript{,}\footnote{For the computation of the phase shift it is useful to have in mind the inverse formula
\begin{equation}
\epsilon_{\bf{a},\ell}^{(1,m)}=  \frac{s^2 i}{2^{15} \pi^3} \int_0^{-s} dt \frac{1}{\sqrt{s}\sqrt{s+t}} \left(1-z^2\right)^{\frac{5}{2}} \frac{C_{\ell}^{\frac{5}{2}}(z)}{C_{\ell}^{\frac{5}{2}}(1)} \left( s^2\times \mathcal{V}_{\bf{a}}^{(1,m)} \right).
\end{equation}
}
\begin{equation}
\epsilon_{\bf{a},\ell}^{(1,0)}(s)=  -\frac{\pi^2}{2}\,\frac{s^2}{(\ell+1)_{4}}
\begin{pmatrix}
 -6  \\
 -2  \\
 -2  \\
 -2 \\
 1  \\
 \hline
 -3  \\
0  \\
\end{pmatrix}.
\end{equation}
Note that the phase shift takes the same form as the AdS$_5 \times$S$^3$ anomalous dimensions, c.f. \eqref{eq:gamma10_singlet}.
As we mentioned already, this striking similarity is to be expected and is related to the existence of the 8-dimensional hidden conformal symmetry.\footnote{This was originally noticed in \cite{Caron-Huot:2018kta} in AdS$_5 \times$S$^5$ and it was one of the first hints that tree-level four-point supergravity amplitudes in AdS$_5 \times$S$^5$ are secretly functions of 10-dimensional variables.}

Let us now consider string corrections. Recall that polynomials in the Mandelstam variables have finite spin support, bounded by the degree of the polynomial in t, thus in general the associated string-corrected partial-wave coefficients are weighted sums of Kronecker deltas:
\begin{equation}\label{flatalphapartial}
\epsilon_{\bf{a},\ell}^{(1,m\geq 2)}(s)=\sum_{\ell'=\{0,1\}}^{m-2-\tfrac{1\pm(-1)^{m+1}}{2}}  a_{\ell'} \delta_{\ell,\ell'}, \qquad \text{ at order} \,\,\ap^{(m+2)}\,,
\end{equation}
where $a_i$ are numbers and the sum runs from $\ell'=0$ ($\ell'=1$) to $m-2-\tfrac{1\pm(-1)^{m+1}}{2}$ for symmetric (antisymmetric) representations.\footnote{This is analogous to AdS amplitudes, where polynomiality in the Mellin variables ensures spin truncation in the conformal block expansion.}

\paragraph{Order $g_s \ap^4$:} At this order we find
\begin{equation}
\mathcal{V}_{\bf{a}}^{(1,2)}=
     256   \zeta_2 \pi^5 \begin{pmatrix}
\frac{15}{2} \\
\frac{7}{2}  \\
-1  \\
 -1  \\
 \frac{1}{2} \\
 \hline
 0  \\
 0  \\
\end{pmatrix},\qquad
\epsilon_{\bf{a},\ell}^{(1,2)}(s)=   \frac{\pi^2}{120}  \zeta_2  \delta_{\ell, 0} s^4 \begin{pmatrix}
\frac{15}{2} \\
\frac{7}{2}  \\
-1  \\
 -1  \\
 \frac{1}{2} \\
 \hline
 0  \\
 0  \\
\end{pmatrix}.
\end{equation}

\paragraph{Order $g_s \ap^5$:} Here we have
\begin{equation}
 \mathcal{V}_{\bf{a}}^{(1,3)}=
   256  \zeta_3 \pi^5 \begin{pmatrix}
 3 s \\
   s\\
  s \\
  s \\
 -\frac{1}{2}  s \\
 \hline
 \frac{3}{2} (t-u) \\
 0  \\
\end{pmatrix}, \qquad
\epsilon_{\bf{a},\ell}^{(1,3)}(s)= 
 \frac{\pi^2}{120} \zeta_3  s^5 \begin{pmatrix}
3 \delta_{\ell, 0}\\
 \delta_{\ell, 0}\\
\delta_{\ell, 0} \\
 \delta_{\ell, 0} \\
-\frac{1}{2} \delta_{\ell, 0} \\
\hline
 \frac{3}{14} \delta_{\ell, 1}  \\
 0  \\
\end{pmatrix},
\end{equation}
where note that now there is a non-zero antisymmetric component which has support on $\ell=1$.

\paragraph{Order $g_s \ap^6$:} Finally, at this order we get
\begin{equation}
\mathcal{V}_{\bf{a}}^{(1,4)}=
   256  \pi^5 \frac{\zeta_2^2}{20} \begin{pmatrix}
 81(t^2 +u^2)+ 99 t u  \\
  37(t^2 +u^2)+ 43 t u  \\
 -8(t^2 +u^2)-2 t u  \\
 -8(t^2 +u^2) -2 t u \\
 4(t^2 +u^2)+  t u \\
\hline
 -9 s (t-u) \\
 0  \\
\end{pmatrix},
\end{equation}
and, correspondingly,
\begin{equation}
\epsilon_{\bf{a},\ell}^{(1,4)}(s)= 
   \frac{\pi^2 }{4800 } \zeta_2^2 s^6  \begin{pmatrix}
135 \delta_{\ell, 0}+ \delta_{\ell, 2}\\
  \frac{425}{7} \delta_{\ell, 0}+ \frac{31}{63} \delta_{\ell, 2} \\
 -10\delta_{\ell, 0} - \frac{2}{9} \delta_{\ell, 2} \\
 -10\delta_{\ell, 0} - \frac{2}{9} \delta_{\ell, 2} \\
 5 \delta_{\ell, 0}+ \frac{1}{9} \delta_{\ell, 2} \\
 \hline
 - \frac{18}{7} \delta_{\ell, 1}  \\
0  \\
\end{pmatrix}.
\end{equation}
Note that the symmetric channels have support on $\ell=0,2$, as they should, since they are polynomials of degree 2 in $t$.

With the phase-shifts at hand we can now compute the one-loop discontinuity. The imaginary part of the one-loop amplitude is given by
\begin{equation}\label{ImV}
s^2 \times \text{Im}[\mathcal{V}_{\bf{a}}^{(2,m\geq 2)}] = \frac{4}{s^{\frac{d-4}{2}}}\sum_{\ell} \sum_{\substack{m',m''\\m'+m''=m}} \left( \epsilon_{\bf{a},\ell}^{(1,m')}\epsilon_{\bf{a},\ell}^{(1,m'')} P_{\ell}^{(5/2)}(z) \right  ), \qquad z=1+\frac{2 t}{s}\,.
\end{equation}
Note that, except for the field-theory one-loop amplitude, the sum truncates order by order in $\ap$, therefore it is immediate to find an explicit expression for the discontinuity of the amplitude.
Moreover, we have factored out an $s^2$ from the above equation, since as explained around \eqref{flatfourstring}, this should be identified with $\mathcal{I}$.

For example, at $\alpha'^6$, equation \eqref{ImV} yields:
\begin{equation}
s^2 \times \text{Im}[ \mathcal{V}_{\bf{a}}^{(2,2)}(s,t)]=  \frac{8}{s^{2}}\epsilon_{\mathbf{a},0}^{(1,0)}\epsilon_{\mathbf{a},0}^{(1,2)} P_{0}^{(5/2)}(z) = s^2 \times  \frac{16}{3} \zeta_2 \pi^7 s^2\begin{pmatrix} 90\\ 14\\ -4\\ -4\\ -1\\\hline 0 \\ 0 \end{pmatrix},
\end{equation}
which is nothing but the imaginary part of \eqref{flatspacelim2}.
Analogously, plugging in the relevant phase shifts, it is easy to check that $\text{Im}[\mathcal{V}_{\bf{a}}^{(2,3)}]$, $\text{Im}[\mathcal{V}_{\bf{a}}''^{(2,4)}]$, $\text{Im}[\mathcal{V}_{\bf{a}}''^{(2,4)}]$ are in agreement with equations \eqref{flatspacelim3}, \eqref{flatspacelim41} and \eqref{flatspacelim42}, respectively.

\section{Conclusions}\label{sec:final}
In this paper, we initiated the study of genus-one open string amplitudes in AdS$_5\times$S$^3$.
The CFT dual to this F-theory construction is a $USp(2N)$ theory with flavour group $SO(8)$, with (the scalar superpartners) of the gluons being dual to half-BPS operators of the form \eqref{contractedO2}.
The main goal of this work was the construction of the leading discontinuity of one-loop correlators at the first three orders in $1/\lambda$, corresponding to the discontinuity of genus-one open string amplitudes in AdS$_5\times$S$^3$. The theory is known to possess a hidden 8-dimensional conformal symmetry in the field-theory limit. While the symmetry is generally broken by string corrections, they still obey a 8-dimensional principle which is encoded in their OPE data. As we described, the set of exchanged double-trace operators is in fact truncated according to their effective 8-dimensional spin $\ell_{8d}$.

This understanding of the structure of the spectrum then allowed us to drastically simplify our one-loop bootstrap program. In particular, for the first three orders -- with the exception of the symmetric channels at order $\lambda^{-2}$ where there are two 8d spin exchanged -- the fact that there is only one 8d spin which is exchanged implies that the one-loop leading log is related to the tree-level discontinuity via the action of the differential operator $\dfour$.

We then bootstrapped the full amplitude in two different representations: Mellin- and position space. In Mellin space, the amplitudes are given by polynomials multiplying digamma functions, whose infinite sequence of single poles is in correspondence with the dimensions of exchanged double-trace operators. In position space, the results can be nicely re-organised in terms of a simple pre-correlator. The full correlator is then obtained from the latter by the action of $(\dfour)^2$. Lastly, as a consistency check of our results, we analysed the flat-space limit of the Mellin amplitudes. In particular, we verified that these are consistent with the $s$-channel discontinuity of the 8d genus-one open string amplitude in flat space, which we explicitly reconstructed order by order in $\ap$ via a partial-wave expansion in 8-dimensional Gegenbauer polynomials.

Looking ahead, we list a few open questions and further directions which we believe are worth exploring:
\begin{itemize}

\item  In this work, we have only considered the correlator with minimal external charges. However, correlators with arbitrary external charges are an important part for carrying the bootstrap program at higher loops, and moreover they can unveil properties which would otherwise not be visible.  One example is the higher-dimensional symmetry itself, which allows to relate all Kaluza-Klein correlators to a single seed, which is precisely correlator with minimal charges.
The structure of these higher-charge correlators will most likely follow similar patterns to those found in \cite{Drummond:2020uni,Aprile:2022tzr} in the context of AdS$_5\times$S$^5$. In particular, the Mellin amplitude will exhibit additional single poles in correspondence with the so-called window and below-window regions, as expected from the structure of the OPE.

\item When considering the position space representation, we noticed that the operator $(\dfour)^2$ can be pulled out. The analogous property is not at all obvious in Mellin space, where $\dfour$ acts as a complicated shift operator on the Mellin variables. It would be interesting to compute the Mellin amplitude of the pre-correlator, and explore if that object has any physical significance.

\item Another interesting follow-up question concerns higher genera. The two-loop field theory correlator has recently been constructed in \cite{Huang:2023oxf}, with the differential operator $\dfour$ playing again a crucial role in their position space ansatz. It would be interesting to explore whether a similar approach is applicable to string corrections at two-loop order. However, the information contained in the leading log is not expected to fix the entire correlator, and therefore additional constraints, such as predictions for the flat-space limit, might need to be supplemented.

\item When reviewing the tree-level amplitudes in Section \ref{sec:tree_level_review}, we pointed out that the $\lambda^{-3/2}$ correction admits the rewriting \eqref{eq:M13_c}. This is a consequence of the BCJ relations being satisfied by $\mathcal{M}^{(1,3)}(1234)$ -- a non-trivial property which is not expected by any generic term in the low-energy expansion. As a direct consequence, also the one-loop correction $\mathcal{M}^{(2,3)}$ features a special colour structure, described in Section \ref{sec:result_M23}. This is in complete analogy with the one-loop field theory term, see \cite{Alday:2021ajh}. In both cases, the colour factors recombine into seemingly gauge-group independent structures, i.e. they become expressible in terms of (products of) structure constants $f^{IJK}$ alone. For the $\lambda^{-3/2}$ correction, this is somewhat surprising given that the methods of \cite{Behan:2023fqq}, which derived $\mathcal{M}^{(1,3)}(1234)$ in the first place, heavily relied on the gauge group being $SO(8)$. It would therefore be interesting to consider tree-level correlators in theories with different gauge groups, and further investigate the potential universality of the $\lambda^{-3/2}$ correction. 

On general grounds, given the knowledge of such tree-level correlators, the methods presented in this paper can then be straightforwardly applied to those other cases, and it is clear from our construction that any group dependence at tree-level will then propagate to one-loop level. The converse, however, is not true in general, since adding up the crossing orientations of the one-loop $s$-channel term is generically a group-dependent operation. It would be instructive to investigate this reasoning in detail with other examples at hand.
\end{itemize}
To conclude, we recall that the 8-dimensional organising principle which these correlators obey -- both at tree- and one-loop level -- is reminiscent of very similar findings in the context of other AdS$\times$S backgrounds. This strongly suggests that the hidden symmetry is the lead actor in this play. What precisely the role of this actor is, and what exactly his parts are, remains a beautiful open question. We believe that an answer to these questions will be of concrete help in the computation of these four-point correlators at all orders in $1/N$ and $1/\lambda$.

\emph{You may say I'm a dreamer, but I'm not the only one.}

\section*{Acknowledgments}
We want to thank Ross Glew for initial collaboration on this project, and Yu-tin Huang, Hikaru Kawai, and Piotr Tourkine for helpful discussions. We are also grateful to Francesco Aprile and Xinan Zhou for comments on the draft. 
MS would like to thank the ICISE institute in Quy
Nhon for hospitality during the early stages of this project.
HP acknowledges support from the ERC Starting Grant 853507, and from the FWO grant ``G094523N -- Holografie en Supersymmetrische Lokalisatie.''
MS is supported by the Ministry of Science and Technology
(MOST) through the grant 110-2112-M-002-006-.

\appendix
\section{$SO(8)$ crossing matrices}\label{app:colour}
The $t$- and $u$-channel crossing matrices for the gauge group $SO(8)$ are given by
\begin{align}\label{eq:t_crossing_mat}
	F_t = \left(
	\begin{array}{ccccccc}
	 \frac{1}{28} & \frac{5}{4} & \frac{5}{4} & \frac{5}{4} & \frac{75}{7} & 1 & \frac{25}{2} \\
	 \frac{1}{28} & \frac{7}{12} & -\frac{5}{12} & -\frac{5}{12} & \frac{5}{7} & \frac{1}{3} & -\frac{5}{6} \\
	 \frac{1}{28} & -\frac{5}{12} & \frac{7}{12} & -\frac{5}{12} & \frac{5}{7} & \frac{1}{3} & -\frac{5}{6} \\
	 \frac{1}{28} & -\frac{5}{12} & -\frac{5}{12} & \frac{7}{12} & \frac{5}{7} & \frac{1}{3} & -\frac{5}{6} \\
	 \frac{1}{28} & \frac{1}{12} & \frac{1}{12} & \frac{1}{12} & \frac{3}{14} & -\frac{1}{6} & -\frac{1}{3} \\
	 \frac{1}{28} & \frac{5}{12} & \frac{5}{12} & \frac{5}{12} & -\frac{25}{14} & \frac{1}{2} & 0 \\
	 \frac{1}{28} & -\frac{1}{12} & -\frac{1}{12} & -\frac{1}{12} & -\frac{2}{7} & 0 & \frac{1}{2} \\
	\end{array}
	\right),
\end{align}
and
\begin{align}\label{eq:u_crossing_mat}
	F_u = \left(
	\begin{array}{ccccccc}
	 \frac{1}{28} & \frac{5}{4} & \frac{5}{4} & \frac{5}{4} & \frac{75}{7} & -1 & -\frac{25}{2} \\
	 \frac{1}{28} & \frac{7}{12} & -\frac{5}{12} & -\frac{5}{12} & \frac{5}{7} & -\frac{1}{3} & \frac{5}{6} \\
	 \frac{1}{28} & -\frac{5}{12} & \frac{7}{12} & -\frac{5}{12} & \frac{5}{7} & -\frac{1}{3} & \frac{5}{6} \\
	 \frac{1}{28} & -\frac{5}{12} & -\frac{5}{12} & \frac{7}{12} & \frac{5}{7} & -\frac{1}{3} & \frac{5}{6} \\
	 \frac{1}{28} & \frac{1}{12} & \frac{1}{12} & \frac{1}{12} & \frac{3}{14} & \frac{1}{6} & \frac{1}{3} \\
	 -\frac{1}{28} & -\frac{5}{12} & -\frac{5}{12} & -\frac{5}{12} & \frac{25}{14} & \frac{1}{2} & 0 \\
	 -\frac{1}{28} & \frac{1}{12} & \frac{1}{12} & \frac{1}{12} & \frac{2}{7} & 0 & \frac{1}{2} \\
	\end{array}
	\right).
\end{align}
Note that the above crossing matrices (together with the $s$-channel matrix given in \eqref{eq:s_crossing_mat}) have the correct properties under multiplication, i.e.
\begin{align}
	F_t^2=F_u^2=F_s^2=\mathbb{I}\,,\qquad F_tF_uF_t=F_uF_tF_u=F_s\,.
\end{align}  

\section{Correlators with arbitrary external KK modes}\label{app:corrgencharges}
In this appendix we collect some useful information for correlators with arbitrary external KK modes. In fact, as explained in the main text, the unmixing procedure we carried makes use of the knowledge of the $\langle ppqq \rangle$ family of correlators.

We start with the definition of reduced correlator, whose generalisation to arbitrary charges reads, in our conventions\footnote{We are following the conventions of \cite{Drummond:2022dxd}.}
\begin{align}
	G_{\vec{p}}^{I_1 I_2 I_3 I_4}(x_i,\eta_i,\bar\eta_i) = G_{0,\vec{p}}^{I_1 I_2 I_3 I_4}(x_i,\eta_i,\bar\eta_i) +\mathcal{P}_{\vec{p}} \, \mathcal{I}\,\mathcal{H}_{\vec{p}}^{I_1 I_2 I_3 I_4}(U,V;y,\bar{y})\,,
\end{align}
where the kinematic factors are then given by
\begin{align}
\mathcal{P}_{\vec{p}}  = \frac{g_{12}^{k_s} g_{14}^{k_t}  g_{24}^{k_u}  \big( g_{13} g_{24} \big)^{p_3} }{\langle \bar{\eta}_1 \bar{\eta}_3\rangle^2 \langle \bar{\eta}_2 \bar{\eta}_4\rangle^2}\,, \quad \mathcal{I} = (x - y)(\bar{x}-y)\,,
\end{align}
with
\begin{align}
k_s \!=\! \frac{ p_1+p_2-p_3-p_4}{2},\,\,\, k_t\!=\! \frac{p_1+p_4-p_2-p_3}{2},\,\,\, k_u \!=\! \frac{ p_2+p_4-p_3-p_1}{2}.\notag
\end{align}

In these conventions, the Mellin transform of the reduced correlator reads
\begin{equation}
\mathcal{H}_{\vec{p}}^{I_1 I_2 I_3 I_4} = \int ds \, dt \sum_{\tilde{s},\tilde{t}} \Gamma_{\otimes} U^s V^t U^{\tilde{s}}V^{\tilde{t}} \mathcal{M}_{\vec{p}}^{I_1 I_2 I_3 I_4}(s,t,\tilde{s},\tilde{t})\,,
\end{equation}
where the stripped-off gamma functions are  
\begin{equation}
\Gamma_{\otimes}=\frac{\Gamma[-s]\Gamma[-s+k_s]\Gamma[-t]\Gamma[-t+k_t]\Gamma[-u]\Gamma[-u+k_u]}{\Gamma[\tilde{s}+1]\Gamma[\tilde{s}+k_s+1]\Gamma[\tilde{t}+1]\Gamma[\tilde{t}+k_t+1]\Gamma[\tilde{u}+1]\Gamma[\tilde{u}+k_u+1]}\,,
\end{equation}
and the AdS Mellin variables and the S sphere variables satisfy the on-shell constraints:
\begin{equation}
s+t+u=-p_3-1\,, \qquad \tilde{s}+\tilde{t}+\tilde{u}=p_3-2\,.
\end{equation}
Note that the choice to single out $p_3$ is purely conventional and depends on the propagator $\mathcal{P}_{\vec{p}}$.
Finally, due to the gamma functions in the denominator, the sum over $\tilde{s},\tilde{t},\tilde{u}$ lies in the triangle:
\begin{equation}
T := \{\tilde{s}\geq \max(0,-k_s),\, \, \tilde{t} \geq 0,\, \, \tilde{u} \geq 0 \}.
\end{equation}

\subsection{Tree-level amplitudes for arbitrary KK correlators}
In this subsection we collect results for arbitrary KK correlators at the first few orders in $\lambda^{-\frac{1}{2}}$.
The field-theory correlator, computed in \cite{Alday:2021odx}, reads, in these conventions \cite{Drummond:2022dxd}:
\begin{equation}
\mathcal{M}^{(1,0)}=-\frac{2}{({\bf s}+1)({\bf t}+1)} \Tr [1234]+\texttt{crossing}\,,
\end{equation}
where the bold-face variables are defined via
\begin{equation}
{\bf s}=s + \tilde{s}, \qquad {\bf t}=t + \tilde{t}\,, \qquad {\bf s}+{\bf t}+{\bf u} =-3\,.
\end{equation}
As an aside, note that the correlator for general charges can be obtained from the correlator with lowest charges by covariantising the Mellin variables into bold-face variables: $s \rightarrow {\bf s},\,  t \rightarrow {\bf t}$.
This is another way of saying that the tree-level correlator enjoys a hidden 8-dimensional conformal symmetry.

String corrections, which are higher derivative corrections are in general polynomials in the variables $s,t,\tilde{s},\tilde{t},p_i$.
At order $\lambda^{-1}$ the polynomial is degree 0 in $s,t,\tilde{s},\tilde{t}$ and takes the form
\begin{equation}
\mathcal{M}^{(1,2)}(1234) = 2^5 \zeta_2 (\Sigma-2)_{2}\,.
\end{equation}
At $\lambda^{-\frac{3}{2}}$ we can combine the results of \cite{Glew:2023wik,Behan:2023fqq} to get the correlator for arbitrary charges. The result of \cite{Glew:2023wik} reads
\begin{equation}
\mathcal{M}^{(1,3)} (1234)= -2^7 \zeta_3 \Big(\mathcal{M}_{1}^s+\mathcal{M}_{1}^t+a_1 (\Sigma-2)_2 \Big),
\end{equation} 
with
\begin{align}
\begin{split}
\mathcal{M}_{1}^s&=(\Sigma-2)_3 {\bf s} - 3 (\Sigma-2)_2 {\check{s}}\,,\\
\mathcal{M}_{1}^t&=(\Sigma-2)_3 {\bf t} - 3 (\Sigma-2)_2 {\check{t}}\,,
\end{split}
\end{align}
and similarly for the other colour-ordered amplitudes.
We also defined shifted variables via
\begin{align}
\begin{split}
& \hat{s}=s+\frac{1}{2}(p_3+p_4)\,, \qquad  \hat{t}=t+\frac{1}{2}(p_2+p_3) \,,\\
& \check{s}={\tilde{s}}-\frac{1}{2}(p_3+p_4)\,, \qquad  \check{t}={\tilde{t}}-\frac{1}{2}(p_2+p_3)\,,
\end{split}
\end{align}
which is a useful definition because it makes crossing symmetry manifest when considering generic charges\footnote{Note that the bold-face variables remain unchanged under the shift.}.
Localisation fixes the remaining ambiguity to be \cite{Behan:2023fqq}
\begin{equation}
a_1=-4\,.
\end{equation}
Finally, the amplitude at order $\lambda^{-2}$ contains 4 ambiguities and reads \cite{Glew:2023wik}:
\begin{equation}
\begin{split}
{\cal M}^{(1,4)} (1234) &= 512\Big[\frac{\pi^4 }{6!}(7{\cal M}_2^s+7{\cal M}_2^t+{\cal M}_2^u)+ a_{2}\,(\Sigma-2)_2 + b_{1}({\cal M}_1^s+{\cal M}_1^t) \\
&\quad\qquad + e_{1} ({\cal M}_{2,\text{amb}}^s+{\cal M}_{2,\text{amb}}^t) + f_{1} {\cal M}_{2,\text{amb}}^u\Big].
\end{split}
\end{equation}
where
\begin{equation}
{\cal M}_2^s=\left(\Sigma-2 \right)_4{\bf{s}}^2- \left(\Sigma-2 \right)_3 {\bf{s}}(8{\check{s}}+ \Sigma +1)+ \left(\Sigma-2 \right)_2 \Big(12{\check{s}}^2-\frac{3}{8} P +12{\check{s}}+\frac{3}{2}\Sigma \Big).
\end{equation}
and analogously for ${\cal M}_2^t,{\cal M}_2^u$.
Here we have defined
\begin{equation}
P \equiv p_1^2+p_2^2+p_3^2+p_4^2\,.
\end{equation}
The ambiguities $e_1,f_1$ are parametrised by the s-type amplitude
\begin{equation}
{\cal M}_{2,\text{amb}}^s=(\Sigma-2)_3 {\bf s}+\frac{3}{14}  (\Sigma-2)_2 \big(p_1 p_2 +p_3 p_4+ 2\Sigma-2\Sigma^2-4\check{s}\Sigma - 2\check{s}\big),
\end{equation}
and its crossing versions ${\cal M}_{2,\text{amb}}^{t,u}$ defined analogously, and $b_1$ is parametrised by the $s$-type amplitude
\begin{equation}
{\cal M}_1^s=(\Sigma-2)_3 {\bf s} -3  (\Sigma-2)_2 \check{s}\,,
\end{equation}
and its crossing version ${\cal M}_{1}^{t,u}$ . Note that ${\cal M}_1^s+{\cal M}_1^t$ is nothing but the amplitude at order $\lambda^{-\frac{3}{2}}$, up to a shift in the constant term.
As we mentioned already, one could use the localisation constraint at this order to fix one of the ambiguities \cite{Behan:2023fqq}.

\section{Tree-level unmixing in string theory}
\label{sec:treeleveunmix}
In this section we give some more details on the unmixing for the family of $\langle ppqq \rangle$ correlators.

\subsection{Long blocks for generic charges}\label{app:long_blocks}
Let us first re-introduce back the $p_i$ dependence in the long blocks. For general charges, these read\footnote{The letter $a$, which together with $b$, labels the R-symmetry representation should not be confused with  $\mathbf{a}$ which instead stands for the SO(8) irrep.}
\begin{equation}
\mathbb{L}_{\vec{\tau}}=\mathcal{P}_{\vec{p}}\,(x-y)(\bar{x}-y)\biggl( \frac{\tilde{U}}{U} \biggr)^{p_3} \mathcal{B}_{\tau,l}(x,\bar{x})\, \mathcal{B}_{b,a}^{\text{int}}(y,\bar{y})\,,
\end{equation}
where
\begin{align}
\begin{split}
	\mathcal{B}_{\tau,l}(x, \bar{x}) &=  \frac{(-1)^l}{(x-\bar{x})U^{\tfrac{p_{43}}{2}}} \Bigl( \mathcal{F}_{\tfrac{\tau}{2}+1+l}^+(x)\mathcal{F}_{\tfrac{\tau}{2}}^+(\bar{x})-\mathcal{F}_{\tfrac{\tau}{2}}^+(x)\mathcal{F}_{\tfrac{\tau}{2}+1+l}^+(\bar{x}) \Bigr),\\
	\mathcal{B}_{b,a}^{\text{int}}(y,\bar{y}) &= \frac{1}{\tilde{U}^{2-\tfrac{p_{43}}{2}}} \mathcal{F}_{-\tfrac{b}{2}-a}^-(y)\mathcal{F}_{-\tfrac{b}{2}}^-(\bar{y})\,,
\end{split}
\end{align}
with
\begin{equation}
\mathcal{F}_{h}^{\pm}(x)= x^h \!\!~_2F_1 \left[h \mp \tfrac{p_{12}}{2},h \mp \tfrac{p_{43}}{2}; 2 h; x \right].
\end{equation}
Note that, unlike the case when $p_i=2$, we also have to deal with the $SU(2)_L \times SU(2)_R$ decomposition. This is achieved by expanding the correlator in terms of the $SU(2)_L \times SU(2)_R$ Jacobi polynomials.
The latter are the product of two $SU(2)$ spherical harmonics, one corresponding to the R-symmetry group $SU(2)_R$ and the other corresponding to the flavour group $SU(2)_L$. Following analogous conventions to those for the long blocks of $\mathcal{N}=4$ SYM, we label them with two numbers $a,b$, which can be viewed as the analogues of twist and spin on the sphere.

Finally, it is worth noticing that the internal blocks are not invariant under $y \leftrightarrow \bar{y}$ exchange. As a consequence, the decomposition is extended to spherical harmonics with label $a<0$. In particular, for given charges $p_i$, we decompose a function in spherical harmonics labelled by the two quantum numbers $[a,b]$. The values of $a$ run over the following set:
\begin{equation}
 -\kappa_{\vec{p}} \leq a \leq \kappa_{\vec{p}}\,,
\end{equation}
where
\begin{equation}
\kappa_{\vec{p}}=\frac{\min(p_1+p_2,p_3+p_4)-p_{43}-4}{2}\,,
\end{equation}
is the so-called degree of extremality and $p_{43}= p_4-p_3$.
For a fixed value of $a$, the quantum number $b$ runs over the set
\begin{equation}
  - \min(a,0) \leq \frac{b-p_{43}}{2} \leq (\kappa_{\vec{p}}-a+ \min(a,0) )\,.
\end{equation}

\subsection{Tree-level unmixing}
Having introduced the necessary tools, we now want to show how to perform an explicit computation of the anomalous dimensions. The goal is to verify the claim we made that the non-zero anomalous dimensions are controlled by the 8-dimensional effective spin. This is ultimately the reason why the leading discontinuity can be obtained by acting with $\dfour$ on the tree-level discontinuity.

Let us first review the double-trace spectrum in AdS$_5\times$S$^3$.
For any given quantum numbers $\vec{\tau}=(\tau,b,l,a)$, the number of double-trace operators exchanged of the form
\begin{equation}
P_\mathbf{a}^{I_1I_2} \mathcal{O}_p^{I_1} \square^{\tfrac{\tau-p-q}{2}} \partial_l \mathcal{O}_q^{I_2}\big|_{[a,b]}\,,
\end{equation}
where $P_\mathbf{a}^{I_1I_2}$ is an appropriate projector which projects onto a symmetric or antisymmetric representation,
is equal to the number of points $(p,q)$ filling the rectangle \cite{Drummond:2022dxd}\footnote{An analogous rectangle was first introduced in the context of supergraviton amplitudes in AdS$_5\times$S$^5$, where it describes exchanged double-trace operators in strongly coupled $\mathcal{N}=4$ SYM \cite{Aprile:2018efk}.}
\begin{align}\label{ir53}
{R}_{\vec{\tau}}:=  \biggl\{(p,q):  &\begin{array}{l}
	p=i+|a|+1+r\\q=i+a+1+b-r\end{array},  \quad	\begin{array}{l}
	i=1,\ldots,(t-1)\\ r=0,\ldots,(\mu -1)\end{array}
	\biggr\},
\end{align}
where 
\begin{align}\label{multiplicity1}
t\equiv \frac{(\tau-b)}{2}- \frac{(a+|a|)}{2}\,,\qquad
\mu \equiv   \left\{\begin{array}{ll}
\bigl\lfloor{\frac{b+a-|a|+2}2}\bigr\rfloor \quad &a+l \text{ even}\,,\\[.2cm]
\bigl\lfloor{\frac{b+a-|a|+1}2}\bigr\rfloor \quad &a+l \text{ odd}\,.
\end{array}\right.
\end{align}

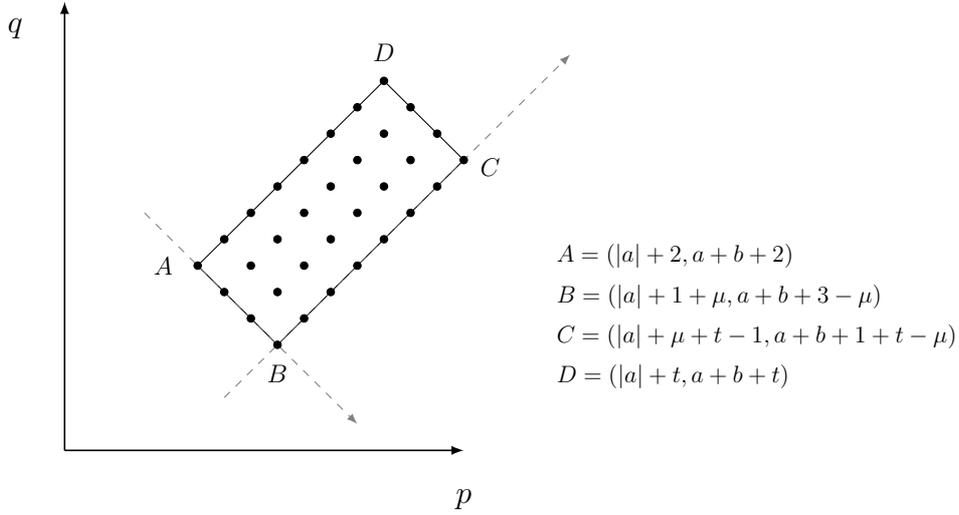
\begin{figure} 
    \centering
\be
\begin{tikzpicture}[scale=.7]
\def\prop{.5}
\def\shifthor{\prop*2}
\def\ptuno{(\prop*2-\shifthor,\prop*8)}
\def\ptdue{(\prop*5-\shifthor,\prop*5)}
\def\pttree{(\prop*9-\shifthor,\prop*15)}
\def\ptquattro{(\prop*12-\shifthor,\prop*12)}
%
%
\draw[-latex, line width=.6pt]		(-2.5,  0.5)    --  (5,0.5) ;
\node[scale=1] (oxxy) at 	(5,1)  {};
\node[scale=1] [below of=oxxy] {$p$};
\draw[-latex, line width=.6pt]		(-2.5,  0.5)    --  (-2.5,9);
\node[scale=1] (oxyy) at 	(-2,8.5)  {};
\node[scale=1] [left of= oxyy] {$q$};
%
\draw[] 								\ptuno -- \ptdue;
\draw[black]							\ptuno --\pttree;
\draw[black]							\ptdue --\ptquattro;
\draw[]								\pttree--\ptquattro;
\draw[-latex,gray, dashed]					(\prop*0-\shifthor,\prop*10) --(\prop*8-\shifthor,\prop*2);
\draw[-latex,gray, dashed]					(\prop*3-\shifthor,\prop*3) --(\prop*16-\shifthor,\prop*16);
%
%
\foreach \indeyc in {0,1,2,3}
\foreach \indexc  in {2,...,9}
\filldraw   					 (\prop*\indexc+\prop*\indeyc-\shifthor, \prop*6+\prop*\indexc-\prop*\indeyc)   	circle (.07);
%
%
\node[scale=.8] (puntouno) at (\prop*3-\shifthor,\prop*8) {};
\node[scale=.8]  [left of=puntouno] {$A$};   
\node[scale=.8] (puntodue) at (\prop*5-\shifthor,\prop*4+1.1) {};
\node[scale=.8] [below of=puntodue]  {$B$}; 
\node[scale=.8] (puntoquattro) at (\prop*13-\shifthor,\prop*14) {};
\node[scale=.8] [below of=puntoquattro] {$C$};
\node[scale=.8] (puntotre) at (\prop*9-\shifthor,\prop*13+.4) {};
\node[scale=.8] [above of=puntotre] {$D$}; 
%
\node[scale=.75] (legend) at (10.5,3) {$\begin{array}{l}  
													\displaystyle A=(|a|+2,a+b+2) \\[.1cm]
													\displaystyle B=(|a|+1+\mu,a+b+3-\mu) \\[.1cm]
													\displaystyle C=(|a|+\mu+t-1,a+b+1+t-\mu) \\[.1cm]
													\displaystyle D=(|a|+t,a+b+t) \\[.1 cm] \end{array}$  };
\end{tikzpicture}
\notag\vspace{-0.6cm}
\ee
\caption{A rectangle of degenerate operators with $\mu=4, t=9$.}
\label{fig:rectads5s3}
\end{figure}

The number of points in the rectangle $ {R}_{\vec{\tau}}$ is $d=\mu(t-1)$.
Figure \ref{fig:rectads5s3} shows an example with $\mu=4, t=9$.
Note that in the singlet $\mu=1$, and the number of points is $\frac{\tau}{2}-2\equiv n-1$, as mentioned in the main text.

Note the appearance of absolute values for $a$. This is a consequence of the fact that the theory is not symmetric under $y\leftrightarrow \bar{y}$ exchange, differently from $\mathcal{N}=4$ SYM.

Note that, for some values of the quantum numbers, the rectangle $R_{\vec{\tau}}$ can degenerate to a line.  
When $\mu=1$ the rectangle collapses to a line with $+45^{\circ}$ orientation; when $\tau=2a+b+4$, with $\mu>1$, which corresponds to the first available twist for the rep $[ab]$, the rectangle also collapses to a line, this time with $-45^{\circ}$ orientation.
Then, as the twist increases the rectangle opens up in the plane.

This representation is useful because at tree-level, operators on the same vertical line remain degenerate. In fact, by solving the $1/N$ OPE equations
\begin{align}
\begin{split}
\mathbf{C}^{(0,0)}{\mathbf{C}^{(0,0)}}^T &= \mathbf{M}^{(0,0)}\,,\\
\mathbf{C}^{(0,0)} \gamma^{(1,0)}{\mathbf{C}^{(0,0)}}^T &= \mathbf{M}^{(1,0)}\,,
\end{split}
\end{align}
one finds that the resulting anomalous dimensions only depend on $p$, but not $q$:
\begin{equation}
	\gamma_{\mathbf{a},p}^{(1,0)}= v_\mathbf{a}\,\frac{\delta_{\tau,l}^{(4)}}{(\ell_{8d}^{\pm}+1)_4}\,,
\end{equation}
which is the generalisation of \eqref{eq:gamma10_singlet} to generic $SU(2)_L \times SU(2)_R$  representations, and we recall that the colour vector $v_\mathbf{a}$ is given by $v_\mathbf{a}=\{-6,-2,-2,-2,1,-3,0\}$. In fact, all that changed is a slight modification in the 8-dimensional effective spin,
\begin{equation}\label{8dspingen}
	\ell_{8d}~ \mapsto ~\ell_{8d}^\pm = \ell+|a| +2(p-2) + \frac{1\mp (-1)^{\ell+a}}{2}\,,
\end{equation}
which now differs for symmetric ($\ell_{8d}^+$) and antisymmetric ($\ell_{8d}^-$) irreps of the flavour group. Moreover, note that $\ell_{8d}^{\pm}$ depends also depends on $a$, but not $b$. For the singlet channel, $[a,b]=[0,0]$, we recover \eqref{8dspin}.\footnote{Recall that in the singlet $\ell \in 2\mathbb{N}$ ($\ell \in 2\mathbb{N}+1$) for symmetric (antisymmetric) irreps, thus $\frac{1\mp (-1)^{\ell+a}}{2}=0$ and $\ell_{8d}^+=\ell_{8d}^-$\, .}

When string corrections are turned on, the left-over degeneracy is sequentially lifted. 
As mentioned in the main text, it turns out that at each order in $\lambda^{-\frac{m}{2}}$ the operators which acquire a correction to their anomalous dimension are governed by the formula
\begin{equation}
	O(\lambda^{-\frac{m}{2}}):\qquad \ell_{8d}^\pm \leq m-2-\frac{1\pm(-1)^{m+1}}{2}\,.
\end{equation}
In correspondence with the above bound, the CFT data is found to satisfy the inequality:
\begin{equation}
\begin{cases}
& \gamma_{\mathbf{a},p}^{(1,m)}=0\,,   \\
& 	 C_{\tilde{p}\tilde{q} \mathcal{S}_{pq}}^{(0,m)}=0\,,	 
\end{cases}
 \qquad \qquad \text{for}\quad \ell_{8d}^\pm > m-2-\frac{1\pm(-1)^{m+1}}{2}\,.
\end{equation}

\begin{figure}
    \centering
\be
\begin{tikzpicture}[scale=.6]
\def\prop{.5}
\def\shifthor{\prop*2}
\def\ptuno{(\prop*2-\shifthor,\prop*8)}
\def\ptdue{(\prop*5-\shifthor,\prop*5)}
\def\pttree{(\prop*9-\shifthor,\prop*15)}
\def\ptquattro{(\prop*12-\shifthor,\prop*12)}
%
\draw[-latex, line width=.6pt]		(-2.5,  0.5)    --  (5,0.5) ;
\node[scale=1] (oxxy) at 	(5,1)  {};
\node[scale=1] [below of=oxxy] {$p$};
%
\draw[-latex, line width=.6pt]		(-2.5,  0.5)    --  (-2.5,9);
\node[scale=1] (oxyy) at 	(-2,8.5)  {};
\node[scale=1] [left of= oxyy] {$q$};
%
\draw[] 								\ptuno -- \ptdue;
\draw[black]							\ptuno --\pttree;
\draw[black]							\ptdue --\ptquattro;
\draw[]								\pttree--\ptquattro;
\draw[-latex,gray, dashed]					(\prop*0-\shifthor,\prop*10) --(\prop*8-\shifthor,\prop*2);
\draw[-latex,gray, dashed]					(\prop*3-\shifthor,\prop*3) --(\prop*16-\shifthor,\prop*16);
%
\foreach \indeyc in {0,1,2,3}
\foreach \indexc  in {2,...,9}
\filldraw   					 (\prop*\indexc+\prop*\indeyc-\shifthor, \prop*6+\prop*\indexc-\prop*\indeyc)   	circle (.07);
%
\draw[red,thick]  (0,4) circle (.3);
\node[scale=1] (legend) at (11,6) {$\begin{array}{l}  
													\displaystyle \textcolor{red}{\ell_{8d}=0} \end{array}$  };			
\end{tikzpicture}
\notag\vspace{-0.6cm}
\ee
\caption{The only operator acquiring a $\lambda$-corrected anomalous dimension at  $\lambda^{-1}$  is circled in red. }
\label{fig:rank1}
\end{figure}
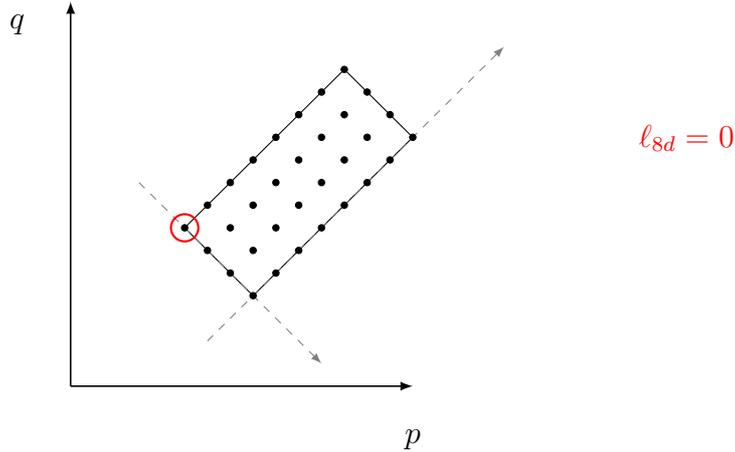

The conjecture can be checked by solving the OPE equations for arbitrary values of the quantum numbers.
For example, at order $\lambda^{-1}$ these read
\begin{align}
\begin{split}
\mathbf{C}^{(0,0)}{\mathbf{C}^{(0,2)}}^T+\mathbf{C}^{(0,2)}{\mathbf{C}^{(0,0)}}^T  &= 0\,,\\
\mathbf{C}^{(0,0)} \gamma^{(1,2)}{\mathbf{C}^{(0,0)}}^T+\mathbf{C}^{(0,0)} \gamma^{(1,0)}{\mathbf{C}^{(0,2)}}^T+\mathbf{C}^{(0,2)} \gamma^{(1,0)}{\mathbf{C}^{(0,0)}}^T &= \mathbf{M}^{(1,2)}\,.
\end{split}
\end{align}
By solving the above equations for many values of twist and spin we find that in fact  \emph{the only non-zero anomalous dimension is the one labeled by $p=2$} (left-most corner of the rectangle in Figure \ref{fig:rank1}) with $\ell=0$, in agreement with the bound \eqref{8dspingen}. The explicit form of the anomalous dimensions is not really needed, its simplicity however deserves some space:
\begin{equation}\label{eq:gamma12}
\gamma_{\mathbf{a},22}^{(1,2)}= -\frac{2\zeta_2}{15}\begin{pmatrix}
15\\7\\-2\\-2\\1 \\\hline 0\\ 0
\end{pmatrix}
\big(\delta_{\tau,\ell=0}\big)^2 \equiv -\frac{2\zeta_2}{15}\,\tilde{v}_\mathbf{a}\, \big(\delta_{\tau,\ell=0}\big)^2\,.
\end{equation}
At this order, the corrected three-point functions $\mathbf{C}^{(0,2)}=0$ vanish.

At order $\lambda^{-\frac{3}{2}}$  the situation is very similar, except that now we have both symmetric and antisymmetric reps. By solving the OPE equations at this order, i.e.
\begin{align}
\begin{split}
\mathbf{C}^{(0,0)}{\mathbf{C}^{(0,3)}}^T+\mathbf{C}^{(0,3)}{\mathbf{C}^{(0,0)}}^T  &= 0 \,,\\
\mathbf{C}^{(0,0)} \eta^{(1,3)}{\mathbf{C}^{(0,0)}}^T+\mathbf{C}^{(0,0)} \gamma^{(1,0)}{\mathbf{C}^{(0,3)}}^T+\mathbf{C}^{(0,3)} \gamma^{(1,0)}{\mathbf{C}^{(0,0)}}^T &= \mathbf{M}^{(1,3)}\,.
\end{split}
\end{align}
one finds again that the bound \eqref{8dspingen} is satisfied, i.e. there is only one operator turned on with label $p=2$, with quantum numbers $\ell=a=0$ for  symmetry irreps, and $\ell=1,a=0$ or $\ell=0,a=1$ for antisymmetric irreps. The corrected three-point functions $\mathbf{C}^{(0,3)}=0$ also vanish. We will refrain from writing down the formulae for the anomalous dimensions, which can be found in \cite{Santagata:2022hga}.

\begin{figure}
    \centering
\be
\begin{tikzpicture}[scale=.6]
\def\prop{.5}
\def\shifthor{\prop*2}
\def\ptuno{(\prop*2-\shifthor,\prop*8)}
\def\ptdue{(\prop*5-\shifthor,\prop*5)}
\def\pttree{(\prop*9-\shifthor,\prop*15)}
\def\ptquattro{(\prop*12-\shifthor,\prop*12)}
\draw[-latex, line width=.6pt]		(-2.5,  0.5)    --  (5,0.5) ;
\node[scale=1] (oxxy) at 	(5,1)  {};
\node[scale=1] [below of=oxxy] {$p$};
\draw[-latex, line width=.6pt]		(-2.5,  0.5)    --  (-2.5,9);
\node[scale=1] (oxyy) at 	(-2,8.5)  {};
\node[scale=1] [left of= oxyy] {$q$};
%
\draw[] 								\ptuno -- \ptdue;
\draw[black]							\ptuno --\pttree;
\draw[black]							\ptdue --\ptquattro;
\draw[]								\pttree--\ptquattro;
\draw[-latex,gray, dashed]					(\prop*0-\shifthor,\prop*10) --(\prop*8-\shifthor,\prop*2);
\draw[-latex,gray, dashed]					(\prop*3-\shifthor,\prop*3) --(\prop*16-\shifthor,\prop*16);
%
%
\foreach \indeyc in {0,1,2,3}
\foreach \indexc  in {2,...,9}
\filldraw   					 (\prop*\indexc+\prop*\indeyc-\shifthor, \prop*6+\prop*\indexc-\prop*\indeyc)   	circle (.07);
%
\draw[red,thick]  (0,4) circle (.25);
\draw[blue,thick]  (.5,4) ellipse (.2 and .9);
\node[scale=1] (legend) at (11,6) {$\begin{array}{l}  
													\displaystyle \textcolor{red}{\ell_{8d}^+=0,2} \\ 
\textcolor{blue}{\ell_{8d}^+=2}
 \end{array}$  };
\end{tikzpicture}
\notag\vspace{-0.6cm}
\ee
\caption{The anomalous dimensions turned on at $\lambda^{-2}$. The one in the left-most corner (red) can have both 8-dimensional spin 0 and 2, depending on the value of $|a|,\ell$. At this order there are also new type of operators turned on (circled in blue) with 8-dimensional spin 2, i.e. $a=\ell=0,p=3$.}
\label{fig:rank3}
\end{figure}
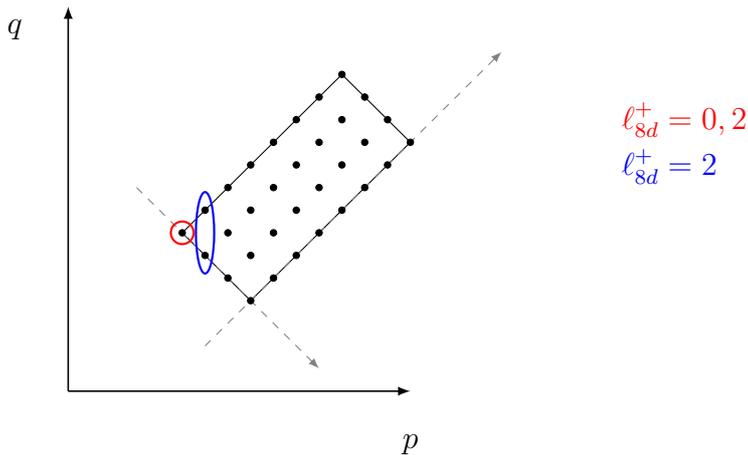

Finally, let us conclude with $\lambda^{-2}$. 
The case of the antisymmetric channels is completely analogous to $\lambda^{-\frac{3}{2}}$, and we only have one anomalous dimension, namely  the one with $p=2$ and $\ell=1,a=0$ or $\ell=0,a=1$.
On the other hand, in the case of symmetric amplitude the situation is slightly more complicated as there are in general multiple operators turned on. Anomalous dimensions and three-point functions are found by solving the equations (remember $\mathbf{C}^{(0,2)}=0$)
\begin{align}
\begin{split}
\mathbf{C}^{(0,0)}{\mathbf{C}^{(0,4)}}^T+\mathbf{C}^{(0,4)}{\mathbf{C}^{(0,0)}}^T  &= 0\,, \\
\mathbf{C}^{(0,0)} \gamma^{(1,4)}{\mathbf{C}^{(0,0)}}^T+\mathbf{C}^{(0,0)} \gamma^{(1,0)}{\mathbf{C}^{(0,4)}}^T+\mathbf{C}^{(0,4)} \gamma^{(1,0)}{\mathbf{C}^{(0,0)}}^T &= \mathbf{M}^{(1,4)}\,.
\end{split}
\end{align}
When $\ell=a=0$, one finds three-anomalous dimensions turned on and $\mathbf{C}^{(0,4)}\big|_{\ell=0} \neq 0$.
Lastly, for values of $a,\ell$ with $|a|+\ell=2$, one again finds only the $p=2$ anomalous dimension turned on, in agreement with \eqref{8dspingen}. Both situations are depicted in Figure \ref{fig:rank3}.

\bibliographystyle{JHEP}
\bibliography{new_bib2}
\end{document}